\documentclass[journal,comsoc]{IEEEtran}

\usepackage[T1]{fontenc}% optional T1 font encoding
\usepackage{xcolor}
\usepackage{fancyhdr}
\usepackage{tgschola}
\usepackage{lastpage}
\usepackage{amsthm,amssymb,amsmath,mathtools,fixmath}
\usepackage{amsbsy}
\usepackage{moreverb}
\usepackage{amsfonts}
\usepackage{graphicx}
\graphicspath{{SimulationFigures/}}
\usepackage{tikz}
\usepackage{listings}
\usepackage{color}
\usepackage{latexsym}
\usepackage{subfigure}
\usepackage{animate}
\colorlet{linkequation}{blue}
\usepackage[linesnumbered,ruled]{algorithm2e}
\usepackage{soul}
\usepackage[T1]{fontenc}
\usepackage[utf8]{inputenc}
\usepackage{tabularx,ragged2e,booktabs}
\usepackage{tabu}
\usepackage{dsfont}
\usepackage{setspace}
\usepackage{enumitem}
\usepackage[utf8]{inputenc}
\usepackage[english]{babel}
\definecolor{Green}{rgb}{0.0, 0.4, 0.0}
\definecolor{babyblue}{rgb}{0.54, 0.81, 0.94}

\newtheorem*{definition1}{Definition 1}
\newtheorem*{definition2}{Definition 2}
\newtheorem*{definition3}{Definition 3}
\newtheorem*{definition4}{Definition 4}

\usepackage{lipsum}% http://ctan.org/pkg/lipsum

\usepackage[colorlinks,bookmarksopen,bookmarksnumbered,citecolor=blue,urlcolor=black]{hyperref}
\hypersetup{linkcolor=blue}

\makeatletter
\def\BState{\State\hskip-\ALG@thistlm}
\makeatother

\usepackage{cite}

\ifCLASSINFOpdf
\else
\fi
\usepackage{amsmath}
\usepackage[cmintegrals]{newtxmath}
\begin{document}
\title{\fontsize{23}{30}\selectfont Distributed User Association in B5G Networks Using Early Acceptance Matching Games}

\author{\normalsize Alireza~Alizadeh,~\IEEEmembership{\normalsize Student Member,~IEEE,} and
        \normalsize Mai~Vu,~\IEEEmembership{\normalsize Senior~Member,~IEEE}
}

\maketitle

\begin{abstract}
We study distributed user association in 5G and beyond millimeter-wave enabled heterogeneous networks using matching theory. 
We propose a novel and efficient distributed matching game, called \textit{early acceptance} (EA), which allows users to apply for association with their ranked-preference base station in a distributed fashion and get accepted as soon as they are in the base station's preference list with available quota. 
Several variants of the EA matching game with preference list updating and reapplying are compared with 
the original and stability-optimal deferred acceptance (DA) matching game, which implements a waiting list at each base station and delays user association until the game finishes. We show that matching stability needs not lead to optimal performance in other metrics such as throughput. Analysis and simulations show that compared to DA, the proposed  EA matching games achieve higher network throughput while exhibiting a significantly faster association process. 
Furthermore, the EA games either playing once or multiple times can reach closely the network utility of a centralized user association while having much lower complexity.
\end{abstract}


\IEEEpeerreviewmaketitle

\section{Introduction} \label{Intro}
\IEEEPARstart{5}{G} and beyond heterogeneous networks (HetNets) will utilize both sub-6 GHz and millimeter wave (mmWave) frequency bands. These networks will be highly dense and composed of different types of base stations (BSs) with different sizes, transmit powers, and capabilities.
In these dense multi-tier networks, finding optimal user associations is a challenging problem, which defines the best possible connections between BSs and user equipment (UEs) to achieve an optimal network performance while satisfying BSs' load constraints. Traditional association method of connecting to the BS with highest SINR may overload certain BSs and no longer work well in a HetNet with different BS transmit powers and under the highly directional and variable mmWave channel conditions. Furthermore, the user association process needs to be fast and efficient to adapt to the low-latency requirements of beyond 5G (B5G) networks.

In a user association problem, the presence or absence of the connection between a UE and a BS is usually indicated by an integer variable with value one or zero, making it an integer optimization. In this paper, we focus on unique user association in which each UE can only be associated with one BS at a time. In B5G HetNets, UEs will be able to work in dual connectivity mode as they are equipped with a multi-mode modem supporting both sub-6 GHz and mmWave bands, allowing the possibility of association to either a macro cell BS (MCBS) or a small cell BS (SCBS).

\subsection{Background and Related Works}
User association for load balancing in LTE HetNets is studied in \cite{Andrews}, where a user association algorithm is introduced by relaxing the unique association constraint and then using a rounding method to obtain unique association coefficients.
In \cite{Caire}, the researchers studied optimal user association in massive MIMO networks and solve the user association problem using Lagrangian duality.
User association in a 60-GHz wireless network is studied in \cite{60GHz}, where the researchers assumed that interference is negligible due to directional steerable antenna arrays and highly attenuated signals with distance at 60-GHz. This assumption, however, becomes inaccurate at other mmWave frequencies considered for cellular use as it is shown that the mmWave systems can transit from noise-limited to interference-limited regimes \cite{Niknam,NoiseOrInterf}. 

The unique user association problem usually results in a complex integer nonlinear programming which is NP-hard in general. Heuristic algorithms have been designed to solve this problem and achieve a near-optimal solution \cite{zalghout2015greedy,TWC}.
These algorithms often require centralized implementation which usually suffers from high computational complexity and relies on a central coordinator to collect the channel state information (CSI) between all BSs and UEs. As a result, practical implementation of these algorithms in mmWave-enabled networks is potentially inefficient given the erratic nature of mmWave channels. 
The authors in \cite{Dist_UA_60} proposed two distributed user association methods for a 60-GHz mmWave network. 
They only considered a simple mmWave channel model based on large-scale CSI and antenna gains, and assumed that the interference is negligible because of highly directional transmissions.

Matching theory has been proposed as a promising low-complexity mathematical framework with interesting applications from labor market to wireless networks. 
In a pioneering work \cite{gale1962college}, Gale and Shapley introduced a matching game, called deferred acceptance (DA) game, to solve the problem of one-to-one and many-to-one matching.
The DA matching game has recently found applications for user association in wireless networks. For example, the DA game is employed for resource management in wireless networks using algorithmic implementations in \cite{gu2015matching}.
A two-tier matching game is proposed for user association based on task offloading to BSs jointly with collaboration among BSs for task transferring to underloaded BSs \cite{UA_MT_MEC}.
Matching algorithms based on the DA game have also been used for user association in the downlink of small cell networks in \cite{semiari2014matching} and for the uplink in \cite{saad2014college}. 

\subsection{Our Contributions}
We propose matching-theory based user association schemes designed by considering system aspects specifically relevant to B5G cellular networks. We consider a mmWave clustered channel model for generating the mmWave links. In the system model, the UEs have dual connectivity and are equipped with both sub-6 GHz and mmWave antenna arrays. The effect of directional transmissions in B5G systems is directly integrated via beamforming transmission and reception in both frequency bands. Moreover, we take into account the dependency between user association and interference which is crucial for beamforming transmission, while previous works either ignore interference or assume it to be independent of user association. Such an assumption is applicable in LTE cellular networks because of omni-directional transmissions, but is no longer suitable for B5G mmWave-enabled networks because of directional transmissions by beamforming at both BSs and UEs.

Existing matching theory user association works are based exclusively on the DA game. One important issue with the DA game is excessive association delay for distributed implementation, as the associations of all UEs are postponed to the end of the game. This is due to the presence of a waiting list at each BS, described in more details in Sec. \ref{DA_subsec}.
In \cite{GLC19}, we proposed a new matching game with lower delay and provided some preliminary results. In this paper, we propose a set of low-delay distributed matching games tailored for user associations in B5G cellular networks.
The main contributions of this paper are:
\begin{itemize}[leftmargin=*]
\item We introduce a distributed user association framework, in which UEs and BSs exchange application and response messages to make decisions, and define relevant metrics to assess its performance. 
We show that stability of a matching game does not directly lead to optimality in other objectives such as the network throughput. 
Instead of stability which is also less relevant in a dynamic network, we consider performance metrics for distributed user association in terms of association delay, user's power consumption in the association process, percentage of unassociated users, and network utility including throughput or spectral efficiency.
\item We propose three novel and purely distributed matching games, called early acceptance (EA), and compare them with the well-known DA matching game. 
Unlike DA which postpones association decisions to the last iteration of the game, our proposed EA games allow the BSs to make their decisions in accepting or rejecting UEs immediately. This approach results in significantly faster association process and at the same time a slightly higher network throughput, and hence presents a better choice for association in fast-varying mmWave systems.
\item Our proposed EA games follow a set of similar rules, but are different in terms of updating the UE/BS preference lists and reapplying to BSs. Numerical results show that the basic EA game without updating or reapplying is the fastest, whereas the EA game with preference list updating and reapplication leads to the highest percentage of associated UEs. Thus, there is a tradeoff between the game simplicity and the number of unassociated UEs.
\item We also propose a multi-game user association algorithm to further improve the network throughput by performing multiple rounds of a matching game. The proposed algorithm requires minimal centralized coordination, and as the number of rounds of the game increases, reaches closely the performance of centralized worst connection swapping (WCS) algorithm introduced in \cite{TWC}. 
\item Our simulations show that the proposed EA games have comparable performance with the DA game in terms of power consumption and signaling overhead, while resulting in a slightly higher network utility and a superior performance in terms of association delay. Considering the fact that the number of UEs to be associated is usually different from the total quota of BSs, we show that our proposed EA games are effective in any loading scenarios (underload, critical load, and overload).
\end{itemize}

\subsection{Notation}
In this paper, scalars and sets are denoted by italic letters (e.g. $x$ or $X$) and calligraphy letters (e.g. $\mathcal{X}$), respectively. Vectors are represented by lowercase boldface letters (e.g. $\mathbf{x}$), and matrices by uppercase boldface letters (e.g. $\mathbf{X}$). Superscript $(.)^T$ and $(.)^*$ represent the transpose operator and the conjugate transpose operator, respectively. $\log(.)$ stands for base-2 logarithm, and big-O notation $\mathcal{O}(.)$ expresses the complexity. $|\mathcal{X}|$ denotes the cardinality of set $\mathcal{X}$.
 $\boldsymbol{I}_N$ is the $N\times N$ identity matrix, and $|\mathbf{X}|$ denotes determinant of matrix $\mathbf{X}$. 

\section{System and User Association Models}\label{Sys_model} 
We study the problem of user association in the downlink of a two-tier HetNet with $B$ macro cell BSs (MCBS) operating at microwave band, $S$ small cell BSs (SCBSs) working at mmWave band, and $K$ UEs. Let $\mathcal{B}$, $\mathcal{S}$, and $\mathcal{J}=\{1, ..., J\}$ denote the respective set of MCBSs, SCBSs, and all BSs with $J=B+S$, and $\mathcal{K}=\{1, ..., K\}$ represents the set of UEs. 
Each BS $j$ has $M_j$ antennas, and each UE $k$ is equipped with two antenna modules: 1) a single antenna for LTE connections at microwave band, and 2) an antenna array with $N_k$ elements for 5G connections at mmWave band. Each UE $k$ aims to receive $n_k$ data streams from its serving BS such that $1\leq n_k\leq N_k$, where the upper inequality indicates that the number of data streams for each UE cannot exceed the number of its  antennas.
We also assume that each UE supports dual-connectivity so that association to either MCBS or SCBS is possible.

\subsection{Microwave and mmWave Channel Models}\label{Ch_Models}
In this subsection, we introduce the microwave and mmWave channel models.
In the microwave band the transmissions are omnidirectional and we use the well-known Gaussian channel model \cite{Telatar}. 
We denote $\mathbf{h}^{\mu\text{W}}$ as the channel vector between a MCBS and a UE where its entries are i.i.d. complex Gaussian random variables with zero-mean and unit variance, i.e., $h^{\mu\text{W}} \sim \mathcal{CN}(0,1)$.
In the mmWave band, the transmissions are highly directional and the Gaussian MIMO channel model no longer applies. We employ the specific clustered mmWave channel model which includes $C$ clusters with $L$ rays per cluster defined as  \cite{3GPP901}, \cite{Nokia}
\begin{align}\label{clustered_ch}
\mathbf{H}^{\text{mmW}}=\frac{1}{\sqrt{CL}}\sum_{c=1}^{C}\sum_{l=1}^{L} \sqrt{\gamma_c}~\mathbf{a}(\phi_{c,l}^{\textrm{UE}},\theta_{c,l}^{\textrm{UE}}) ~\mathbf{a}^*(\phi_{c,l}^{\textrm{BS}},\theta_{c,l}^{\textrm{BS}})
\end{align}
where $\gamma_c$ is the power gain of the $c$th cluster. The parameters $\phi^{\textrm{UE}}$, $\theta^\textrm{UE}$, $\phi^\textrm{BS}$, $\theta^\textrm{BS}$ represent azimuth angle of arrival, elevation angle of arrival, azimuth angle of departure, and elevation angle of departure, respectively. The vector $\mathbf{a}(\phi,\theta)$ is the response vector of a uniform planar array (UPA) which allows 3D beamforming in both the azimuth and elevation directions.
We consider the probability of LoS and NLoS as given in \cite{RapLetter}, and utilize the path loss model for LoS and NLoS links as given in \cite{3GPP901}. 
The numerical results provided in Sec. \ref{Sim_res} are based on these channel models and related parameters.

\subsection{Signal Model}
For tier-1 working at sub-6 GHz band, the effective interfering channel on UE $k$ from MCBS $j\in\mathcal{B}$ serving UE $l$ is defined as
\begin{equation}
h_{k,l,j} = \mathbf{h}^{\mu\text{W}}_{k,j}\mathbf{f}_{l,j}
\end{equation}
where $\mathbf{f}_{l,j}\in\mathbb{C}^{M_j\times 1}$ is the linear precoder (transmit beamforming vector) at MCBS $j$ intended for UE $l$. If $l=k$, this defines the effective channel between MCBS $j$ and UE $k$ as 
$h_{k,j} = \mathbf{h}^{\mu\text{W}}_{k,j}\mathbf{f}_{k,j}$.

Similarly, for tier-2 operating at mmWave band, the effective interfering channel on UE $k$ from SCBS $j\in\mathcal{S}$ serving UE $l$ is defined as
\begin{equation}
\mathbf{H}_{k,l,j} = \mathbf{W}^*_k\mathbf{H}^\text{mmW}_{k,j}\mathbf{F}_{l,j}
\label{H_mmW}
\end{equation}
where $\mathbf{F}_{l,j}\in\mathbb{C}^{M_j\times n_l}$ is the linear precoder at SCBS $j$ intended for UE $l$, and $\mathbf{W}_k\in\mathbb{C}^{N_k \times n_k}$ is the linear combiner (receive beamforming matrix) of UE $k$.
If $l=k$, (\ref{H_mmW}) becomes the effective channel between SCBS $j\in\mathcal{S}$ and UE $k$ which includes both beamforming vectors/matrices at the BS and UE, and can be expressed as $\mathbf{H}_{k,j} = \mathbf{W}^*_k\mathbf{H}^\text{mmW}_{k,j}\mathbf{F}_{k,j}$.
Thus, the received signal at UE $k$ connected to MCBS $j\in\mathcal{B}$ can be written as
\begin{equation}\label{y_k_muW}
y_k^{\mu\text{W}} = \sum_{j\in \mathcal{B}}h_{k,j}s_{k,j} + z_k
\end{equation}
where $s_{k,j}\in \mathbb{C}$ is the data symbol intended for UE $k$ with $\mathbb{E}\lbrack s^*_{k,j}s_{k,j}\rbrack =P_{k,j}$,
and $z_k\in\mathbb{C}$ is the complex additive white Gaussian noise at UE $k$ with $z_k\sim\mathcal{CN}(0,N_0)$, and $N_0$ is the noise power. 
We consider an equal power allocation scheme to split each BS $j$ transmit power ($P_j$) equally among its associated users, i.e., $P_{k,j}=P_j/|\mathcal{K}_j|$.

Similarly, the received signals at UE $k$ connected to SCBS $j\in\mathcal{S}$ is given by
\begin{equation}\label{y_k_mmW}
\mathbf{y}_k^\text{mmW} = \sum_{j\in \mathcal{S}}\mathbf{H}_{k,j}\mathbf{s}_{k,j} + \mathbf{W}^*_k\mathbf{z}_k
\end{equation}
where $\mathbf{s}_{k,j}\in \mathbb{C}^{n_k}$ is the data stream vector for UE $k$ consisting of mutually uncorrelated zero-mean symbols with $\mathbb{E}\lbrack \mathbf{s}^*_{k,j}\mathbf{s}_{k,j}\rbrack =P_{k,j}$,
and $\mathbf{z}_k\in\mathbb{C}^{N_k}$ is the complex additive white Gaussian noise vector at UE $k$.
The presented signal model, developed algorithms and insights for user association in this paper is applicable to all types of channel models, transmit beamforming and receive combining.

\subsection{User Association Model}
\begin{figure}
\centering
\includegraphics[scale=.45]{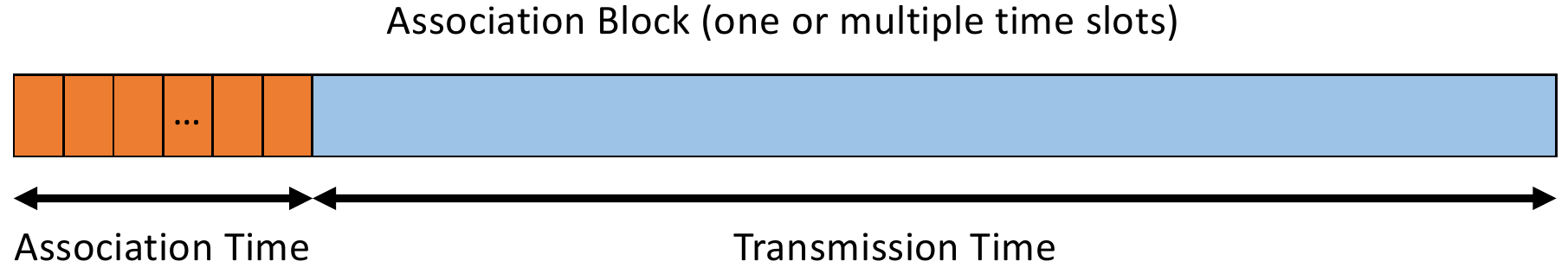}
\caption{Structure of an association block: User association is established in a distributed fashion during the association time interval then is applied to the transmission time interval.}
\label{Ass_block}
\end{figure}

We follow the mmWave-specific user association model introduced in \cite{TWC}. Because of directional beamforming in mmWave systems, the interference structure depends on the user association. Taking into account this dependency is important for mmWave systems where the channels are probabilistic and fast time-varying, and the interference depends on the highly directional connections between BSs and UEs.
We perform user association per a time duration which we call an \textit{association block}, which can span a single or multiple time slots depending on the availability of CSI and is a design choice (Fig. \ref{Ass_block}).

We assume a model where in each association block, the user association process occurs in the association time interval which establishes the UE-BS connections for transmission time interval. The association time interval is further divided into sub-slots for distributed implementation, where during each sub-slot UEs can apply to a BS for association. In a fully distributed algorithm, UE-BS associations can be determined at the end of each sub-slot within the association time interval. This is fundamentally different from a centralized implementation where all user associations are determined at the end of the association time interval. 
In distributed implementation, the duration of the association time interval can vary for each UE's association block, depending on the delay in the association process for that UE.
In the next association block, the user association process needs to be performed again to update  associations according to users' mobility and channel variations.
An association block can represent a single time slot, when both instantaneous (small-scale) and large-scale CSI are available, or can be composed of several consecutive time slots when only large-scale CSI is available. 
Such a choice will lead to a trade-off between user association overhead and resulting network performance. 

We study a \textit{unique user association} problem in which each UE can be served by only one BS in each association block. In this problem, UE-BS associations can be completely determined by an association vector $\mathbold{\beta}$ defined as follows
\begin{equation}\label{Beta_eq}
\mathbold{\beta}=[\beta_1, ..., \beta_K]^T
\end{equation}
and \textit{unique association constraints} can be expressed by
\begin{align}\label{TFA_cons_01}
\sum_{j\in\mathcal{J}} \mathds{1}_{\beta_k=j}&\leq 1, ~\forall k\in \mathcal{K}
\end{align}
where $\beta_k$ represents the index of BS to which user $k$ is associated, and $\mathds{1}_{\beta_k=j}$ is the indicator function such that $\mathds{1}_{\beta_k=j}=1$ if $\beta_k=j$,  and $\mathds{1}_{\beta_k=j}=0$ if $\beta_k\neq j$. All analysis and results in this paper are per association block, and thus we do not consider a time index in our formulations. 

In a HetNet composed of different types of BSs, each BS $j$ can have a different \textit{quota} $q_j$, i.e., the number of UEs it can serve simultaneously.
We define $\mathbf{q}=[q_1,...,q_J]$ as the \textit{quota vector} of BSs, and $\mathcal{K}_j$ as the \textit{activation set} of BS $j$ which represents the set of active UEs in BS $j$, such that $\mathcal{K}_j \subseteq \mathcal{K}$, $|\mathcal{K}_j|\leq q_j$. 
Thus, we can define \textit{load balancing constraints} for BSs as
\begin{align}\label{TFA_cons_02}
\sum_{k\in\mathcal{K}}\mathds{1}_{\beta_k=j} &\leq q_j, ~\forall j\in \mathcal{J}
\end{align}
This set of constraints denotes that the number of UEs served by BS $j$ can not exceed its quota $q_j$. The load balancing constraints allow our formulation to specify each BS’s quota separately. This makes the resulting user association scheme applicable to HetNets where there are different types of BSs with different capabilities.

When UE $k$ is connected to MCBS $j\in\mathcal{B}$, its \textit{instantaneous rate} (in bps/Hz) is
\begin{equation}\label{R_kj_muW}
R_{k,j}^{\mu\text{W}}(\mathbold{\beta}) = \log_2\left(1+ \frac{P_{k,j}h_{k,j} h_{k,j}^*}{v_{k,j}(\mathbold{\beta})}\right)
\end{equation}
where $P_{k,j}$ represents the transmit power from BS $j$ dedicated to UE $k$, and $v_{k,j}$ is the interference plus noise given as 
\begin{align}
v_{k,j}(\mathbold{\beta}) = \sum_{\substack{l\in \mathcal{K}_{j} \\ l\neq k}} P_{l,j}h_{k,l,j}h_{k,l,j}^*+\sum_{\substack{i\in \mathcal{B} \\ i\neq j}} \sum_{\substack{l\in \mathcal{K}_i}} P_{l,i}h_{k,l,i}h_{k,l,i}^*+
N_0 \nonumber
\end{align}
Similarly, the instantaneous rate of UE $k$ connected to SCBS $j\in\mathcal{S}$ is given by
\begin{equation}\label{R_kj_mmW}
R_{k,j}^\text{mmW}(\mathbold{\beta}) = \log_2\Big |\mathbf{I}_{n_k} + \mathbf{V}_{k,j}^{-1}(\mathbold{\beta})P_{k,j}\mathbf{H}_{k,j} \mathbf{H}_{k,j}^*\Big |
\end{equation}
where $\mathbf{V}_{k,j}$ is the interference and noise covariance matrix given as 
\begin{align}
\mathbf{V}_{k,j}(\mathbold{\beta})&=\sum_{\substack{l\in \mathcal{K}_{j} \\ l\neq k}} P_{l,j}\mathbf{H}_{k,l,j}\mathbf{H}_{k,l,j}^*\nonumber \\&+
\sum_{\substack{i\in \mathcal{S} \\ i\neq j}} \sum_{\substack{l\in \mathcal{K}_i}} P_{l,i}\mathbf{H}_{k,l,i}\mathbf{H}_{k,l,i}^* + N_0 \mathbf{W}_k^*\mathbf{W}_k \nonumber
\end{align}
Note that the summations in $v_{k,j}$ and $\mathbf{V}_{k,j}$ are taken over the activation set of BSs, indicating the dependency between interference and user association.

\subsection{Centralized vs. Distributed User Associations}\label{centralized_vs_distributed}
User association is usually studied in the literature in the form of centralized algorithms. Centralized algorithms can reach a near-optimal solution, however, they require a central coordinator, for example, located in a cloud-radio access network (C-RAN), to collect all required CSI and run the user association algorithm. 
In this centralized structure, BSs transmit CSI reference signals via physical downlink control channels (PDCCHs) to enable UEs to estimate the CSI. The CSI is sent back to BSs and then forwarded to C-RAN where the central coordinator is located. 
Since at each iteration of a centralized algorithm, such as the WCS algorithm in \cite{TWC}, user instantaneous rates are updated based on the current association vector, raw CSI and not the SINR must be available to compute these user rates. After collecting all required CSI from the network, the central coordinator runs the user association algorithm to find the best possible connections. The signaling overhead in this ideal centralized structure is usually high due to network densification in B5G HetNets, which requires a significant amount of CSI to be reported, leading to high computational cost and time complexity as the network size increases. 

Distributed user association algorithms have been introduced as low-complexity approaches with a reasonably low convergence time. These algorithms only involve low bit-rate signaling exchanges between UEs and BSs such that association decisions happen in a distributed fashion without a need for a central coordinator.
We assume the UEs exchange messages with the BSs directly and the association decision for different UEs can occur asynchronously at different times. In this paper, we employ matching theory and propose a new matching game to solve the user association problem in a distributed fashion. We also introduce a multi-game matching algorithm to further improve the network performance.  

\section{Matching Theory for Distributed User Association}
\label{MT_for_DUA}
Matching theory attracted the attention of researchers due to its low-complexity and fast convergence time \cite{gu2015matching}. These promising features make matching theory a suitable framework for distributed user association in fast-varying mmWave systems.
User association can be posed as a matching game with two sets of players -- the BSs and the UEs. In this game, each player collects the required information to build a preference list based on its own objective function using local measurements. 
Each user then apply to the BSs based on its preference list, and association decision is made by the BSs individually. 
Thus, no central coordinator is required and user association can be performed fully distributed. This feature makes matching theory an efficient approach for designing a distributed user association in B5G HetNets.

\subsection{User Association Matching Game Concepts}
In the context of matching theory, user association problem can be considered as a college admission game where the BSs with their specific quota represent the colleges and UEs are considered as students. This framework is suitable for user association in a HetNet where BSs may have different quotas and capabilities. In order to formulate our user association as a matching game, we first introduce some basic concepts based on two-sided matching theory \cite{roth1992two}.
\begin{definition1}
A \textit{preference relation} $\succeq_k$ helps UE $k$ to specify the preferred BS between any two BSs $i,j \in \mathcal{J},~i\neq j$ such that
\begin{equation}\label{prf_rlt_K}
j \succeq_k i \Leftrightarrow \Psi_{k,j}^{\text{UE}} \geq \Psi_{k,i}^{\text{UE}} \Leftrightarrow \text{UE}~k~\text{prefers BS}~j~\text{to BS}~i
\end{equation}
where $\Psi^{\text{UE}}_{k,j}$ is the \textit{preference value} between UE $k$ and BS $j$, which can be simply a local measurement at the UE (e.g. SINR).
Similarly, for any two UEs $k,l \in \mathcal{K},~k\neq l$, each BS builds a preference relation $\succeq_j$ such that 
\begin{equation}\label{prf_rlt_J}
k \succeq_j l \Leftrightarrow \Psi_{k,j}^{\text{BS}} \geq \Psi_{l,j}^{\text{BS}} \Leftrightarrow \text{BS}~j~\text{prefers UE}~k~\text{to UE}~l
\end{equation}
\end{definition1}
\begin{definition2}
Based on the preference relations, each UE $k$ (BS $j$) builds its own \textit{preference list} $\mathcal{P}_k^\text{UE}$ ($\mathcal{P}_j^\text{BS}$) over the set of all BSs (UEs) in descending order of preference. 
\end{definition2}
\begin{definition3}
Association vector $\mathbold{\beta}$ defines a \textit{matching relation}\footnote{In this paper we use the terms "association vector" and "matching relation" interchangeably.}, which specifies the association between UEs and BSs and has the following properties
\begin{enumerate}
\item $\beta_k \in \mathcal{J}$ with $k\in\mathcal{K}$;
\item $\beta_k=j$ if and only if $k\in \mathcal{K}_j$.
\end{enumerate}
The second property states that the association vector $\mathbold{\beta}$ is a bilateral matching.
\end{definition3}
\begin{definition4}
A user association \textit{matching game} $\mathcal{G}$ is a game with two sets of players (BSs and UEs) and a set of rules which apply on the input data to obtain a result. The input data of the game are:
\begin{itemize}
\item Preference lists of BSs: $\mathcal{P}_j^\text{BS}, \forall j\in\mathcal{J}$
\item Preference lists of UEs: $\mathcal{P}_k^\text{UE}, \forall k\in\mathcal{K}$
\item BSs' quota: $q_j, \forall j\in\mathcal{J}$
\end{itemize}
and game outcome or result is the association vector $\mathbold{\beta}$. Each particular game is defined by the specific way of building preference lists and its set of rules.
\end{definition4}

\subsection{Building Preference Lists}\label{building_prf_lists}
A preliminary step in a matching game is to build the preference lists of the players. In this subsection, we describe the process of building the preference lists for the UEs and BSs.
These preference lists can be built based on a number of metrics, including users' instantaneous rates, channel norms or UEs' local measurements. 

\subsubsection{Users' instantaneous rates}
Using users' instantaneous rates to build preference lists will require the knowledge of both instantaneous and large-scale CSI. For example, we can use the user instantaneous rate in (\ref{R_kj_muW})-(\ref{R_kj_mmW}) as the objective function for both sides of the game, i.e., 
\begin{equation}\label{R_as_PrfList}
\Psi_{k,j}^{\text{UE}}=\Psi_{k,j}^{\text{BS}}=R_{k,j},~\forall k\in\mathcal{K},j\in\mathcal{J}
\end{equation}

\noindent
In other words, each UE prefers the BS which provides the highest user instantaneous rate, and each BS is willing to serve the UE that can get the highest user instantaneous rate. 
This metric, however, requires frequent update of the preference value which depends on the association of other users and can complicate the process. 
One approach is fixing the association of all other UEs based on an initial association vector for computing the current user's instantaneous rates from all BSs. Alternatively, we can switch the association of the current UE with another UE connected to a BS while computing the instantaneous rate from that BS to the current UE, in order to maintain the BS quota. This switching can be either random or with the weakest UE connected to that BS  as in \cite{TWC}.
\subsubsection{Channel norms}
The preference lists can also be built based on just CSI. As stated earlier, we assume both instantaneous and large-scale CSI available through beamforming CSI estimation techniques. In this case, the preference values can be expressed as the Frobenius norm of a MIMO channel
\begin{equation}
\Psi_{k,j}^{\text{UE}}=\Psi_{k,j}^{\text{BS}}=||\mathbf{H}_{k,j}||_F,~\forall k\in\mathcal{K},j\in\mathcal{J}
\end{equation}

\noindent
where $\mathbf{H}_{k,j}$ is the channel matrix between UE $k$ and BS $j$, and the operator $||.||_F$ represents the Frobenius norm. Using the channel norm, each UE (BS) ranks the BSs (UEs) and builds its own preference list such that the BS (UE) with the strongest channel (highest channel norm) is the most preferred one. The preference lists can also be built based on large-scale CSI alone, which will stay valid for longer durations but may result in a lower overall network utility.
\subsubsection{Local CQI measurements}
A more practical approach to building the preference lists is based on UEs' local measurements. Using reference signal received power information, each UE measures the received SINR from each BS as a ratio of a valid signal to non-valid signals. Then, the UE converts this SINR information to CQI values and reports them to BSs via the PUCCH signaling mechanism \cite{3GPP_PDCCH}. In this approach, the preference values are given by
\begin{equation}
\Psi_{k,j}^{\text{UE}}=\Psi_{k,j}^{\text{BS}}=\text{CQI}_{k,j},~\forall k\in\mathcal{K},j\in\mathcal{J}
\end{equation}
where $\text{CQI}_{k,j}$ represents the channel quality between UE $k$ and BS $j$. Using CQI values, each UE (BS) ranks the BSs (UEs) such that the BS (UE) with the highest CQI is the most preferred one. 
The periodicity of CQI report is a configurable parameter and it could be as fast as every four time slots\footnote{The duration of time slot in 5G NR depends on transmission numerology, and is less than 1ms which is the subframe duration \cite{3GPP_PHY}. Thus, CQI can be reported as fast as every 4ms.} \cite{3GPP_RRC}. 
%\textcolor{red}{CQI values are attractive in the sense that they are simple to measure locally at each UE and can be quickly reported to BSs through uplink control information \cite{3GPP_UCI}.}

\begin{figure}
\centering
  \centering
  \includegraphics[scale=.625]{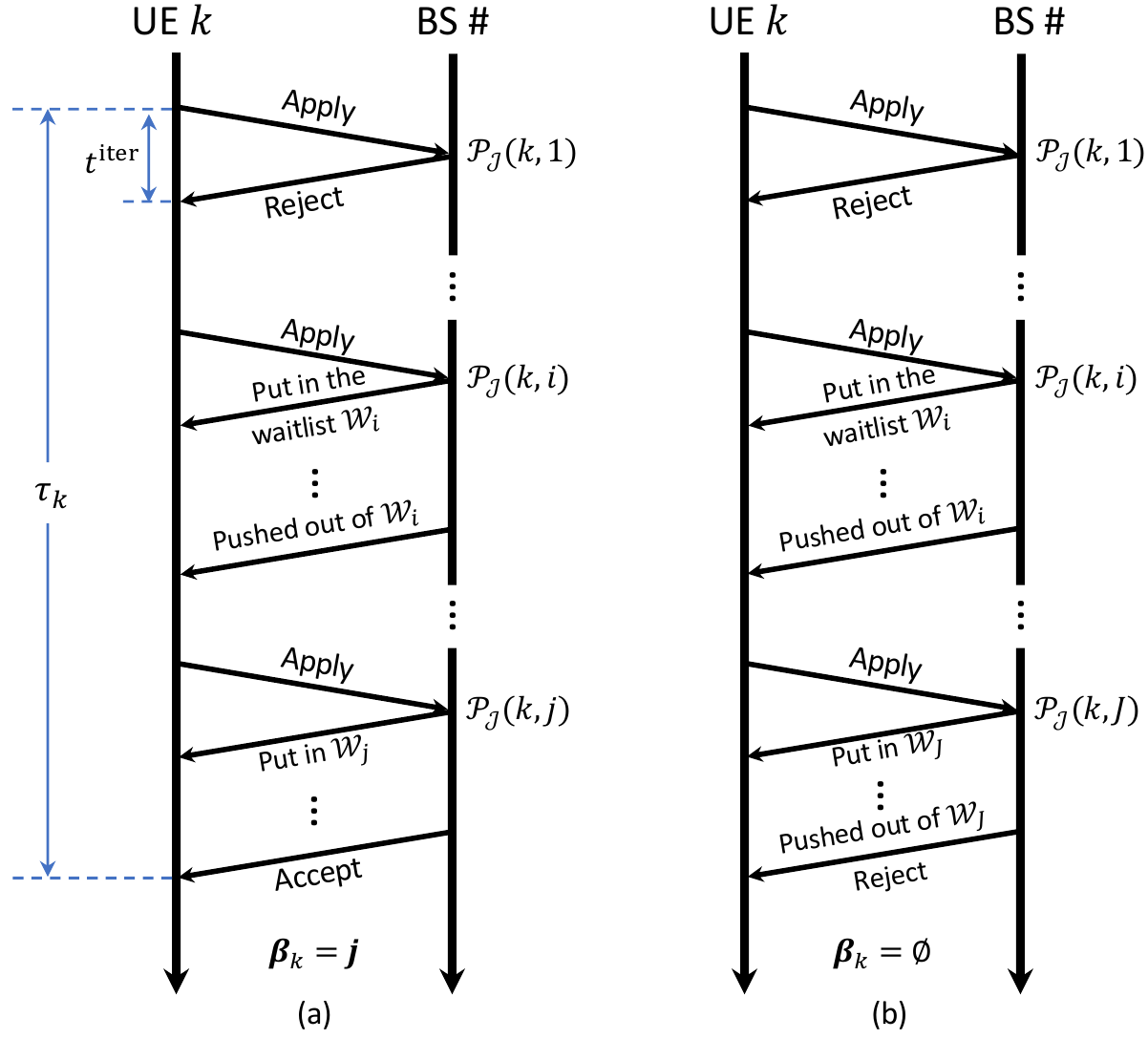}
   \caption{User association process in the DA game. (a) UE $k$ is associated with BS $j$, (b) UE $k$ is pushed out of waiting lists and eventually rejected by all BSs.}
    \label{DA_game_fig}
   % M130_MT_Journal_NumApp_AssDelay.m
\end{figure}

%https://patentimages.storage.googleapis.com/63/4e/b0/1a7d335e0eb082/EP2800412A1.pdf
\subsection{Deferred Acceptance \cite{gale1962college} User Association Matching Game}\label{DA_subsec}
Matching theory dates back to the early 60s when mathematicians Gale and Shapley proposed the now-famous DA matching game, which can be posed as a college admission game and produces optimal and stable results \cite{gale1962college}. Applying to user association, the input data of this game are preference lists of BSs and UEs as well as the quota of BSs, and its result is a matching relation $\mathbold{\beta}$. 
While the DA game can be implemented in a centralized way, in this paper, we focus on its distributed implementation, which does not require a central entity to collect the preference lists of all BSs and UEs and run the game \cite{gu2015matching}.
In what follows, we describe the rules of this game in the context of user association.
\begin{algorithm}[t]\small
\SetAlgoLined
\KwData{$\mathcal{P}_j^\text{BS}$, $\mathcal{P}_k^\text{UE}$, $q_j$, $\forall k\in\mathcal{K},j\in \mathcal{J}$}
\KwResult{Association vector $\mathbold{\beta}=[\beta_1, \beta_2, ..., \beta_K]$}
\textbf{Initialization}: %randomly generate $\mathbold{\beta}^0$ according to BSs' quota\;
Set $m_k=1,~\forall k$, $n=1$, form a rejection set $\mathcal{R}=\{1, 2, ..., K\}$, initialize a set of unassociated UEs $\mathcal{U}=\varnothing$ and the waiting list of each BS $\mathcal{W}_j^0=\varnothing,~\forall j$\;
\While{$\mathcal{R}\neq\varnothing$}{
Each UE $k\in\mathcal{R}$ applies to its $m_k$th preferred BS\;
Each BS $j$ forms its current waiting list $\mathcal{W}_j^{n}$ from its new applicants and its previous waiting list $\mathcal{W}_j^{n-1}$\;
Each BS $j$ keeps the first $q_j$ preferred UEs from $\mathcal{W}_j^{n+1}$, and reject the rest of them\;
\For{$k\in\mathcal{R}$}{
$m_k\leftarrow m_k+1$\;
\If{$m_k>J$}{
Remove UE $k$ from $\mathcal{R}$ and add it to $\mathcal{U}$\;
}
}
$n\leftarrow n+1$\;
}
Form $\mathbold{\beta}$ based on the final waiting lists of BSs $\mathcal{W}_j,~j=1,...,J$.
%\tcp{At the end of matching game, we have $\mathbf{v}_{\textrm{rej}}=\varnothing$.} 
\caption{DA User Association Game (based on \cite{gale1962college})}
\label{DA_Game}
\end{algorithm}

We first define $m_k$ as the \textit{preference index} of UE $k$, $\mathcal{R}$ as the \textit{rejection set} of UEs, $\mathcal{U}$ as the \textit{set of unassociated UEs}, and $\mathcal{W}_j$ as the \textit{waiting list} of BS $j$. Before starting the game, we set the preference index of each UE to one ($m_k=1,~\forall k$), form a rejection set $\mathcal{R}$ including all $K$ UEs, initialize a set of unassociated UEs ($\mathcal{U}=\varnothing$) and the waiting list of each BS ($\mathcal{W}^0_j=\varnothing,~\forall j$). At $n$th iteration of the game, each UE $k$ applies to its $m_k$th preferred BS by sending an application message.
Each BS $j$ then ranks its new applicants together with the UEs in its current waiting list ($\mathcal{W}_j^{n-1}$) based on its preference list, keeps 
the first $q_j$ UEs in its new waiting list $\mathcal{W}_j^n$, and rejects the rest of UEs.
Each BS accordingly sends a response of either rejection or waitlisting to new UE applicants as well as those previously in its waiting list but are now rejected. 
Rejected UEs remain in the rejection set, update their preference index $m_k$, and apply to their next preferred BS in the next iteration.
If $m_k>J$, it means UE $k$ is applied to all BSs and get rejected. Thus, we remove UE $k$ from rejection set $\mathcal{R}$ and add it to the set of unassociated UEs $\mathcal{U}$.
At each new iteration, each BS forms a new waiting list by selecting the top $q_j$ UEs among the new applicants and those on its current waiting list. The game terminates when the rejection set is empty.

In the DA game, the associations are only determined when the game terminates. 
The BSs use their waiting lists to keep the most preferred UEs over all application rounds, and the final waiting lists after the last iteration determine the associations.
This DA user association process is shown in Fig. \ref{DA_game_fig} and
the algorithm for the DA user association game is described in Alg. \ref{DA_Game}.

\subsection{User Association Matching Game Metrics}\label{MG_Metrics}
\subsubsection{Game Stability vs. Other Objectives}
\label{Stability_sec}
In the context of matching theory, stability has been considered as a key performance metric \cite{gale1962college}. Stability means there is no blocking pair, a UE-BS pair $(k,j)\notin\mathbold{\beta}$ where they prefer each other more than their associations under matching relation $\mathbold{\beta}$ \cite{gale1962college}.

Stability in matching is an important criterion as discussed in the original paper by Gale and Shapley \cite{gale1962college}, who showed that DA is optimal in terms of stability. There is, however, no direct implication that an optimal matching in terms of stability is also optimal in terms of another performance objective, including the metric used to build the preference list. We will illustrate this lack of connection via an example. Table \ref{Prf_table} shows the preference lists of the players (in the same format as in \cite{gale1962college}): the first number of each cell is the preference of the Greek-letter player for the Roman-letter player, and the second number is Roman for Greek. Each preference list is built based on the metric values (for example, spectral efficiency in user association) in Table \ref{MetVal_table}.
We can see that the stable assignment here is $(\alpha,A)$ and  $(\beta,B)$ as shown in bold in Table \ref{Prf_table} . The other assignment set of $(\alpha,B)$ and  $(\beta,A)$ is inherently unstable since  $\alpha$ prefers $A$ more than $B$, and $A$ also prefers $\alpha$ more than $\beta$ (per definition of stability as in \cite{gale1962college}). The total metric value for the stable assignment, however, is 5 and is less than the value of the unstable assignment of 6. Thus, optimality in terms of stability does not necessarily result in the highest (optimal) metric values. As such, when applying to a cellular system where a performance metric such as spectral efficiency is of primary interest, a stable association does not necessarily lead to the optimal spectral efficiency. This fact will also be confirmed via simulation results in Sec. \ref{Sim_res}  (see Fig. \ref{SumRate}).

\begin{table}[t]%[!htb]
\setlength{\tabcolsep}{3pt}
\begin{minipage}{.5\linewidth}
\centering
\caption{Preference lists of players}
\label{Prf_table}
\begin{tabular}{cccc}
    \toprule
     & $~~~~A~~~~$& $~~~~B~~~~$\\
    \midrule
    $~~~\alpha~~$ & \textbf{(1,1)} &  (2,1)  \\
    $~~~\beta~~$ & (1,2) & \textbf{(2,2)} \\
    \bottomrule
\end{tabular}
\end{minipage}\hfill
\begin{minipage}{.5\linewidth}
\centering
\caption{Metric values of players}
\label{MetVal_table}
\begin{tabular}{ccc}
    \toprule
     & $~~~~A~~~~$& $~~~~B~~~~$\\
    \midrule
    $~~~\alpha~$ & 4 & 3  \\
    $~~~\beta~$ & 3 & 1 \\
    \bottomrule
\end{tabular}    
\end{minipage}\hfill
\end{table}

Furthermore, stability becomes less relevant in a setting where users' preferences change over time, as is the case of a cellular system. 
Instead of using stability which doesn't apply in a dynamic system where user association can change at every association slot, we define four relevant metrics for distributed user association matching games.
\subsubsection{User Association Delay} This metric represents the amount of time it takes for a UE to get associated with a BS. Denote the time delay for one iteration as $t^\text{iter}$, which is a system parameter and independent of the matching game. Thus, we can compare the association delay of different matching games by comparing their respective time delay until association. 
In a DA game, the association decision is postponed to the last iteration of the game due to the presence of waiting lists. Thus for DA, all UEs have the same association delay ($\tau_{k,\text{DA}}=\tau_\text{DA}$) which is proportional to the total number of iterations $N_\text{DA}^\text{iter}$ of the game as follows
\begin{equation}
\tau_\text{DA}\triangleq N^\text{iter}_\text{DA}t^\text{iter}
\end{equation}
For an EA game, as will be described in Sec. \ref{EA_MG}, the $k^\text{th}$ UE's association delay is
\begin{equation}
\tau_{k,\text{EA}}\triangleq N^\text{appl}_kt^\text{iter}
\end{equation}
where $N_k^\text{appl}$ is the number of applications of UE $k$ until it is associated with a BS. 
Thus, the average association delay for the EA game can be obtained as
\begin{equation}
\tau_{\text{EA}} = \frac{1}{K}\sum_{k\in\mathcal{K}}\tau_{k,\text{EA}}
\end{equation}
We assume that all applicants apply simultaneously at each iteration of the game, thus the time duration for each iteration is the same for every applicant.
\subsubsection{User Power Consumption for Association} In a matching game, each UE applies to its most preferred BSs, until it is accepted by one of them. Each application requires the UE to send a signal to a BS, which is a power-consuming process. 
We assume that each application consumes a specific amount of power ($P^\text{appl}$), which is also a system parameter independent of the matching game. Thus, the power consumption of UE $k$ during the association process can be defined as
\begin{equation}
\gamma_k\triangleq N^\text{appl}_k P^\text{appl}
\end{equation}
Thus, different user power consumption during the association process can be compared by considering the users' number of applications, the lower the number of applications of a UE, the less its power consumption.
\subsubsection{Percentage of Unassociated Users} In the case the number of UEs is more than the total quota of BSs, or some BS are out of range such that the preference lists of certain UEs are shorter than $J$, there may be unassociated UEs at the end of a matching game. We can evaluate the performance of matching games by comparing the percentage of unassociated UEs under these games. In Sec. \ref{Sim_res}, we consider the following scenarios in evaluating the performance of matching games.
\begin{itemize}
\item \textit{Underload}: $K<\sum_{j\in\mathcal{J}} q_j$
\item \textit{Critical load}: $K=\sum_{j\in\mathcal{J}} q_j$
\item \textit{Overload}: $K>\sum_{j\in\mathcal{J}} q_j$
\end{itemize}
\subsubsection{Network Utility Function}
A network utility function can be used to compare the performance of centralized and distributed user association algorithms. In particular, we employ sum-rate utility function to assess the performance of association schemes. 
Defining the instantaneous user throughput vector $\mathbf{r}(\mathbold{\beta})\triangleq (R_{1,\beta_1}, ..., R_{K,\beta_K})$, we can express the sum-rate utility function as
\begin{equation}\label{sum_rate}
U(\mathbf{r}(\mathbold{\beta}))\triangleq \sum_{k\in\mathcal{K}} R_{k,\beta_k}
\end{equation}
where $R_{k,\beta_k}$ is the instantaneous rate of UE $k$ associated with BS $\beta_k$ given in (\ref{R_kj_muW})-(\ref{R_kj_mmW}).

\begin{figure}
  \centering
\includegraphics[scale=0.625]{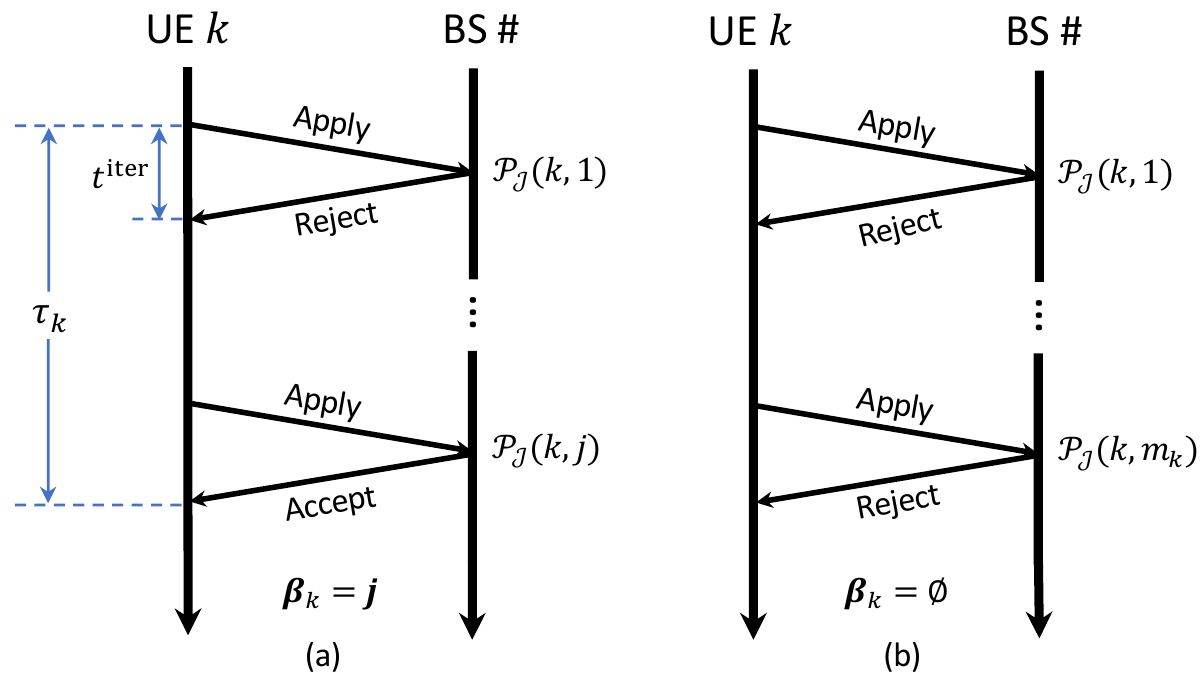}
     \caption{User association process in an EA game. (a) UE $k$ is associated with BS $j$, (b) UE $k$ is not associated as it is rejected at the $m_k$th (last) iteration of the game, with $m_k\geq J$.}
    \label{EA_game_fig}
  % M130_MT_UnderLoad_to_OverLoad.m
% load('M130_MT_UnderLoad_to_OverLoad_100loc_20ch_5BSs_15q_1.mat')
\end{figure}

\section{Proposed Early Acceptance Matching Games}\label{EA_MG} 
The original DA game defers matching decision until the last iteration of the game, and thus is suitable for processes which do not require making decisions in real time. When applied to user association, however, all UEs are kept in BSs' waiting lists until the game terminates. This game can result in an excessive delay for the association process and can be problematic when it comes to user association in fast-varying mmWave systems which require low-latency communications. 
In order to overcome this problem, we propose a set of new matching games, called \textit{early acceptance} (EA) games, to solve the user association problem in B5G HetNets. In an EA game, BSs immediately decide about acceptance or rejection of UEs at each application round. This leads to a significantly faster and more efficient user association process. 
The DA game however, has an advantage over the EA games in terms of stability, but this property is less relevant in the user association context since the preference lists of users change with time. Furthermore, stability needs not lead to optimal throughput, as shown in Sec. \ref{Stability_sec}. Our simulations also confirm that the EA game achieves slightly higher network throughput but at a much lower delay.

\subsection{Proposed Early Acceptance User Association Matching Games}
We introduce three distributed EA matching games which follow a set of similar rules, but are different in terms of updating the UE/BS preference lists and reapplying to BSs. Similar to the DA game, an EA game takes as input data the preference lists of BSs and UEs and quota of BSs, and delivers a matching relation $\mathbold{\beta}$. The initial steps of these games is similar to that of DA which is to set the preference index of all UEs to one ($m_k=1,~\forall k$), form a rejection set $\mathcal{R}$ composed of all $K$ UEs, and initialize a set of unassociated UEs ($\mathcal{U}=\varnothing$).
\subsubsection{EA-Base Game}
This game defines the basic rules of the EA games. At each iteration of the game, each UE $k\in\mathcal{R}$ applies to its $m_k$th preferred BS (say BS $j$) regardless of its available quota. If UE $k$ is among the top $q_j$ UEs in the preference list of BS $j$, it will be immediately accepted by this BS, and the game updates the association vector with $\beta_k=j$. The accepted UEs will be removed from the rejection set $\mathcal{R}$ and any rejected UEs will apply to their next preferred BS in the next iteration. If a UE applies to all the BSs in its preference list and get rejected from all of them, we add the UE to the set of unassociated UEs ($\mathcal{U}$).

This game is the simplest form EA since the UEs and BSs do not update their preference list during the game, and if a UE is rejected by a BS it will not reapply to that BS. These features make the EA-Base game very fast with only a small number of applications ($\leq J$) for each UE. However, at the end of this game some UEs may be unassociated regardless of loading scenarios. 
\subsubsection{EA-PLU (EA with Preference List Updating) Game}
In order to improve the performance of EA-Base, we update the preference lists of UEs and BSs at the end of each iteration. 
When an association happens, the following updates occur: 1) associated UEs are removed from the rejection set and the preference lists of all BSs, and 2) for each new association with BS $j$, its quota is updated as $q_j\leftarrow q_j-1$. When a BS runs out of quota, it informs all the UEs by sending a broadcast message and the UEs remove this BS from their preference lists. As a result, the UEs will no longer apply to that BS. Similar to EA-Base though, there is still no guarantee that all UEs are associated at the end of the EA-PLU game.

\begin{algorithm}[t]\small
\SetAlgoLined
\KwData{$\mathcal{P}_j^\text{BS}$, $\mathcal{P}_k^\text{UE}$, $q_j$, $\forall k\in\mathcal{K},j\in \mathcal{J}$}
\KwResult{Association vector $\mathbold{\beta}=[\beta_1, \beta_2, ..., \beta_K]$ }
\textbf{Initialization}: 
Set $m_k=1,~\forall k$, form a rejection set $\mathcal{R}=\{1, 2, ..., K\}$, and initialize a set of unassociated UEs $\mathcal{U}=\varnothing$\;
\While{$\mathcal{R}\neq\varnothing$}{
Each UE $k\in\mathcal{R}$ applies to its $m_k$th preferred BS (namely BS $j$ with $q_j\neq 0$)\;
\eIf{$k \in \mathcal{P}_j^\text{BS}(1:q_j)$}{
$\beta_k=j$\;
$q_j\leftarrow q_j-1$\;
\If{$q_j=0$}{
Remove BS  $j$ from $\mathcal{P}_k^\text{UE},~\forall k\in\mathcal{K}$\;
   }Remove UE  $k$ from $\mathcal{R}$ and $\mathcal{P}_j^\text{BS},~\forall j\in\mathcal{J}$\;}{
\eIf{$\mathcal{P}_k^\text{UE}\neq\varnothing$}{   
$m_k\leftarrow m_k+1$\;
Keep UE $k$ in $\mathcal{R}$\;}{
Remove UE $k$ from $\mathcal{R}$ and add it to $\mathcal{U}$\;
}
  }
}
\caption{Proposed EA-PLU-RA User Association Game}
\label{EA_Game}
\end{algorithm}

\subsubsection{EA-PLU-RA (EA with Preference List Updating and ReApplying) Game}\label{EA-PLU-RA}
In order to improve the percentage of associated users, we allow each UE to reapply to those BSs from which it has been rejected in previous iterations. 
Recall that if UE is rejected by a BS, it will be kept in the rejection set.
In the following iterations, each UE applies to the next preferred BS in its updated preference list. 
When a UE reaches the end of its updated preference list, it comes back to the beginning and repeats the application process until its updated preference list becomes empty or it is associated. In the case that the updated preference list becomes empty, no reapplication is possible and the UE is removed from the rejection set $\mathcal{R}$ and added to the set of unassociated UEs $\mathcal{U}$.
With this reapplication, when the length of the original preference list of each UE is equal to $J$, no UE is unassociated at the end of the game in underload and critical load cases (see Lemma 2). In case of an incomplete original preference list, UE unassociation is possible. The user association process in the EA-PLU-RA game is depicted in Fig. \ref{EA_game_fig}, and described in Alg. \ref{EA_Game}.

We note that it is practical to update the preference lists of UEs and BSs after each round of applications during an EA matching game. When UE-BS associations occur, the associating BSs update their quota and preference lists. They also inform all other BSs (via backhaul links) and UEs (via a cell broadcast message \cite{3GPP_CB}) to update their preference lists accordingly. In particular, each BS updates its quota and remove associated UEs from its preference list. If a BS runs out of quota, all unassociated UEs remove that BS from their preference lists. This reporting mechanism incurs minimal signaling overhead in practical implementations.

\subsection{Convergence of Matching Games}\label{Conv_MGs}
For a matching game, convergence implies that the game will eventually terminate and produce a matching result. 
Depending on the game and loading scenario, different events can happen at convergence. In the EA-Base and EA-PLU games which do not allow re-application, convergence occurs when the last
unassociated UE(s) applies to its least preferred BS or all BSs run out of quota. In the EA-PLU-RA game, convergence occurs when either all UEs are associated or all BSs are full. We provide four lemmas on the convergence of the proposed EA matching games and obtain their \textit{worst convergence time} (maximum number of iterations). 

\textbf{Lemma 1}. \textit{EA-Base and EA-PLU games always converge and their worst convergence time is $J$ iterations.} 
\begin{proof}
EA-Base and EA-PLU games converge when the last unassociated UE(s) applies to its least preferred BS after having been rejected from all others. Since each UE can only apply once to each BS (reapplying is not allowed), the games terminate in at most $J$ iterations.
\end{proof}

\textbf{Lemma 2}. \textit{For underload and critical load cases, when the length of the original preference list of each UE is equal to the number of BSs $J$, all UEs will be associated at the end of the EA-PLU-RA game.}
\begin{proof}
If the length of the original preference list of each UE is equal to $J$, i.e., $|\mathcal{P}^\text{UE}_k|=J,~\forall k\in\mathcal{K}$, then any rejected UE(s) will have a chance to apply to all BSs and therefore will be associated as long as there is quota left at any of the BSs, which is always the case in underload and critical load scenarios. But if there is a UE with $|\mathcal{P}^\text{UE}_k|<J$, the UE can not apply to all BSs and become unassociated if all BSs in its preference list run out of quota.
\end{proof}

\textbf{Lemma 3}. \textit{The EA-PLU-RA game converges within finite number of iterations.}
\begin{proof}
As stated in Alg. \ref{EA_Game}, the EA-PLU-RA game terminates when the rejection set $\mathcal{R}$ becomes empty. 
According to Lemma 2, for underload and critical load cases when the length of the original preference list of each UE is equal to $J$, all UEs will be associated by the end of the game and the rejection set becomes empty. If there is a UE with a shorter original preference list than $J$, it keeps reapplying to the BSs in its preference list. Once a BS runs out of quota, it will be removed from the preference list of the UE. If all the BSs run out of quota, the preference list of the UE becomes empty and the UE will be moved to set $\mathcal{U}$ (unassociated) and the rejection set becomes empty.
For the overload scenario, the game converges when all BSs run out of quota, which will eventually happens because of preference list update and reapplication. After this point the preference lists of all rejected UEs are empty, and they will be moved from $\mathcal{R}$ to $\mathcal{U}$, causing the game to terminate. 
\end{proof}

Because of reapplication, the worst convergence time of EA-PLU-RA is longer than of the other two EA games and is harder to determine as it depends on the preferences setting. The next lemma provides an upper bound on the worst convergence time of the EA-PLU-RA game.

\textbf{Lemma 4}. \textit{For a network with $J$ BSs and $K$ UEs, an upper bound on the maximum number of iterations for the EA-PLU-RA game under critical load scenario is $N_{K,J}^\text{max} = J(K-J)+\sum_{j=1}^{J}j$.}
\begin{proof}
We prove the bound by reduction. When there are more than $J$ quotas left in the network, then at least one UE must be associated after every $J$ iterations. The worst case is when no association occurs in the first $J-1$ iterations, then there must be at least one UE associated at the $J^\text{th}$ iteration. Using this logic, then when there are $J$ unassociated users left, an upper bound on the maximum number of iterations up to this point is $J(K-J)$. From this point on, if there are $L$ quotas left, then the longest time it takes to get at least one UE associated is after $L$ applications, in the case those quotas belong to different BSs. Thus after at most $L$ applications, the number of unassociated UEs and available quota reduces to $L-1$. Using this reduction, we can obtain an upper bound on the maximum number of iterations for a network with $J$ BSs and $K$ UEs as $N_{K,J}^\text{max} = J(K-J)+\sum_{j=1}^{J}j$.
\end{proof}
The bound in Lemma 4 is conservative and quite loose as indicated by our numerical results, nevertheless it provides a concrete cutoff value on the worst convergence time. The actual worst convergence time is found numerically to be significantly smaller than the bound.

\subsection{Complexity of Matching Games}\label{Comp_MGs}
In this subsection, we analyze the complexity of user association matching games.

1) \textit{Computation complexity of building preference lists}: Prior to starting the game, we need to build the preference lists of BSs and UEs. This process needs to be performed by each player of the game. As mentioned earlier, preference lists can be built based on user's instantaneous rate or some local measurements at the UE. 
In practical scenarios, this information can be measured at a minimal computational complexity using the mechanism described in Sec. \ref{building_prf_lists}. 
In a distributed matching game, each UE can locally measure the received SINR from each BS separately, then compute the instantaneous rate using a single computation. After computing the instantaneous rates, the UE sorts these rates as a mean of ranking the BSs in a descending order to build its preference list. This sorting step results in a computational cost of $\mathcal{O}(J\log(J))$ at each UE. Similarly, we obtain the computational cost of building a preference list at each BS as $\mathcal{O}(K\log(K))$. As a result, the total computational cost for building the preference lists of all UEs and BSs is $\mathcal{O}(JK\log(JK))$.

2) \textit{Game execution complexity}: During a matching game, each UE applies to a BS by sending an application message, and it is notified about the BS decision via a response message.
This process is the same for DA and EA games at the UEs' side, and the number of iterations specifies the execution complexity for each game. However, the execution complexity of each game is different at the BSs' side because the BSs respond to applicants in different ways. 
At each iteration of the DA game, each BS requires to perform a sorting procedure (Step 5) with cost $\mathcal{O}(K\log(K))$, which incurs a total computational cost of order $\mathcal{O}(JK\log(K))$ at each iteration. In the EA games, no such sorting is required by the BSs, and thus they have a much lower execution complexity than the DA game on the BSs' side. 

Our numerical results show that the number of iterations of these games is usually around the number of BSs ($J$). Thus, the total execution complexity of the DA game and EA games are $\mathcal{O}(J^2K\log(K))$ and $\mathcal{O}(J)$, respectively. 
We assume BSs and UEs perform their actions based on their most recently updated preference lists through the reporting mechanism described in Sec. \ref{EA-PLU-RA}. Thus, updating the preference lists in the EA-PLU and EA-PLU-RA games does not increase the execution complexity of these games.
Considering both the computational complexity of building the preference lists and executing the game, the total computational complexity for the DA game and EA games are $\mathcal{O}(J^2K\log(K))$ and $\mathcal{O}(JK\log(JK))$, respectively.
In the complexity analysis of the centralized WCS algorithm in \cite{TWC}, we showed that the total complexity of that algorithm is $\mathcal{O}(M_j^2K^2\log(K))$, which is much higher than the one for distributed matching games since $M_j\gg J$. The computational complexities of the centralized, distributed, and semi-distributed (discussed later in Sec. \ref{Comp_MA}) user association schemes are summarized in Table \ref{Comp_UA_schemes}.
%%%%%%%%%%%%%%%%%%%%%%%%%%%%%%%%%%%%%%%%%%%%
\begin{algorithm}[t]\small
\SetAlgoLined
\KwData{$\mathcal{J}$, $\mathcal{K}$, $\mathbold{q}_j, \forall j\in \mathcal{J}$, Path loss information}
\KwResult{Near-optimal association vector $\mathbold{\beta}^\star$ }
\textbf{Initialization}: \\
- Set the number of games ($N$)\;
- Build initial preference lists of BSs and UEs ($\mathcal{P}_{j}^0$, $\mathcal{P}_{k}^0$, $\forall k,j$) based on channel norms\; 
- Perform a matching game (DA or EA) to obtain initial $\mathbold{\beta}^1$\;
\For{$n=1:N$}{
Calculate $R_{k,j}(\mathbold{\beta}^n),~\forall k, j$\;
Build preference lists $\mathcal{P}_{j}^n$, $\forall  j\in \mathcal{J}$ and $\mathcal{P}_{k}^n$, $\forall k\in\mathcal{K}$\;
Perform a matching game (DA or EA) to obtain $\mathbold{\beta}^{n+1}$\;
}
$\mathbold{\beta}^\star=\mathrm{arg}~\max_{n=1,...,N} ~U(\mathbf{r}(\mathbold{\beta}^n))$.
\caption{\small Multi-Game Matching Algorithm for User Association with Max-throughput}
\label{EA_Alg}
\end{algorithm}

\section{Multi-Game Matching Algorithm}\label{Dist_UA}
The distributed matching games are fast and efficient in terms of delay and power consumption, but due to the distributed nature, they may not reach the performance of centralized algorithms. If we can afford some delay and additional power consumption as well as a minimal signaling exchange, we can further enhance the performance in terms of a network utility. In this section, we introduce a user association optimization problem which aims to maximize a network utility function, then propose a multi-game matching algorithm which requires running multiple rounds of a game and a central entity to keep track of the best association vector. Each game is still run in an entirely distributed fashion, and only the resulting association vector is sent to the central entity for tracking.

Due to the dependency between user association and interference structure of the network, user instantaneous rates or local measurements could change with different associations. Thus, the preference lists may change according to user associations. In particular, at the end of a user association matching game, we obtain an association vector $\mathbold{\beta}$ which specifies the UE-BS connections. Since the user instantaneous rate is a function of $\mathbold{\beta}$, the resulting preference lists at the end of a game round may be different from the original one at the start of the same round, and performing another round of a matching game may produce a better user association in terms of a network utility. In order to keep improving the network performance, we introduce a matching algorithm which plays multiple rounds of a matching game in an iterative manner.
Each round of a game is aimed to maximize the sum-rate as in (\ref{sum_rate}), given the initial association vector obtained from previous round of the game.

\subsection{Multi-Game Matching Algorithm for Max-throughput}
This matching algorithm requires an initial association vector $\mathbold{\beta}^1$ which can be obtained by performing a user association matching game in the initialization procedure. The preference lists of BSs and UEs for this initial game can be built based on channel norms as  described in Sec. \ref{MT_for_DUA}.B. 

At each subsequent iteration of the algorithm, by fixing the associations of all other UEs based on the current association vector $\mathbold{\beta}^n$, each UE computes the instantaneous rate it can get from each BS and reports this rate to the corresponding BS. 
Then, each BS (UE) updates its preference list by ranking all UEs (BSs) based on the computed instantaneous rates. 
Next, a matching game (DA or EA) is performed to obtain the new association vector $\mathbold{\beta}^{n+1}$. This new association vector will be used to establish the preference lists for the next round. 

The algorithm performs the matching game $N$ times, where $N$ is a design parameter. At each time, the game is run in a distributed fashion.
At the end of each game, the BSs report their associations to a central entity, called \textit{best-$\mathbold{\beta}$-tracker}, which computes the utility function and keeps track of the best association vector. As $N$ increases, there is a higher chance of obtaining a better association at the cost of more association delay and higher power consumption at the UEs. The value of $N$ can be determined based on practical delay and power constraints. At the end of the algorithm, the best-$\mathbold{\beta}$-tracker notifies the BSs with the best association vector corresponding to the highest network utility. Although this algorithm requires a central entity to keep track of the best association vector, at each round, the game is performed in a purely distributed manner. This algorithm is described in Alg. \ref{EA_Alg}.

\begin{table}[t]
\centering
   \caption{Computational complexity of centralized, distributed, and semi-distributed user association schemes}
 {\tabulinesep=1.2mm
   \begin{tabu} {|c|c|}
    \hline \small
User Association Scheme & \small Complexity \\
       \hline \scriptsize
       WCS Algorithm (centralized)\cite{TWC}  & \scriptsize $\mathcal{O}\Big(M_j^2K^2\log(K)\Big)$  \\\hline \scriptsize
       DA Game (distributed)\cite{gale1962college}  & \scriptsize $\mathcal{O}\Big(J^2K\log(K)\Big)$ \\\hline \scriptsize
        Proposed EA-PLU-RA Game (distributed)   & \scriptsize $\mathcal{O}\Big(JK\log(JK)\Big)$\\\hline \scriptsize
        Proposed Multi-game DA Alg. (semi-distributed)   & \scriptsize $\mathcal{O}\Big(NJ^2K\log(K)\Big)$\\\hline \scriptsize
        Proposed Multi-game EA Alg. (semi-distributed) & \scriptsize $\mathcal{O}\Big(NJK\log(JK)\Big)$\\\hline
   \end{tabu}}
   \label{Comp_UA_schemes}
\end{table}

\begin{figure*}[t]
\centering
\includegraphics[scale=.31]{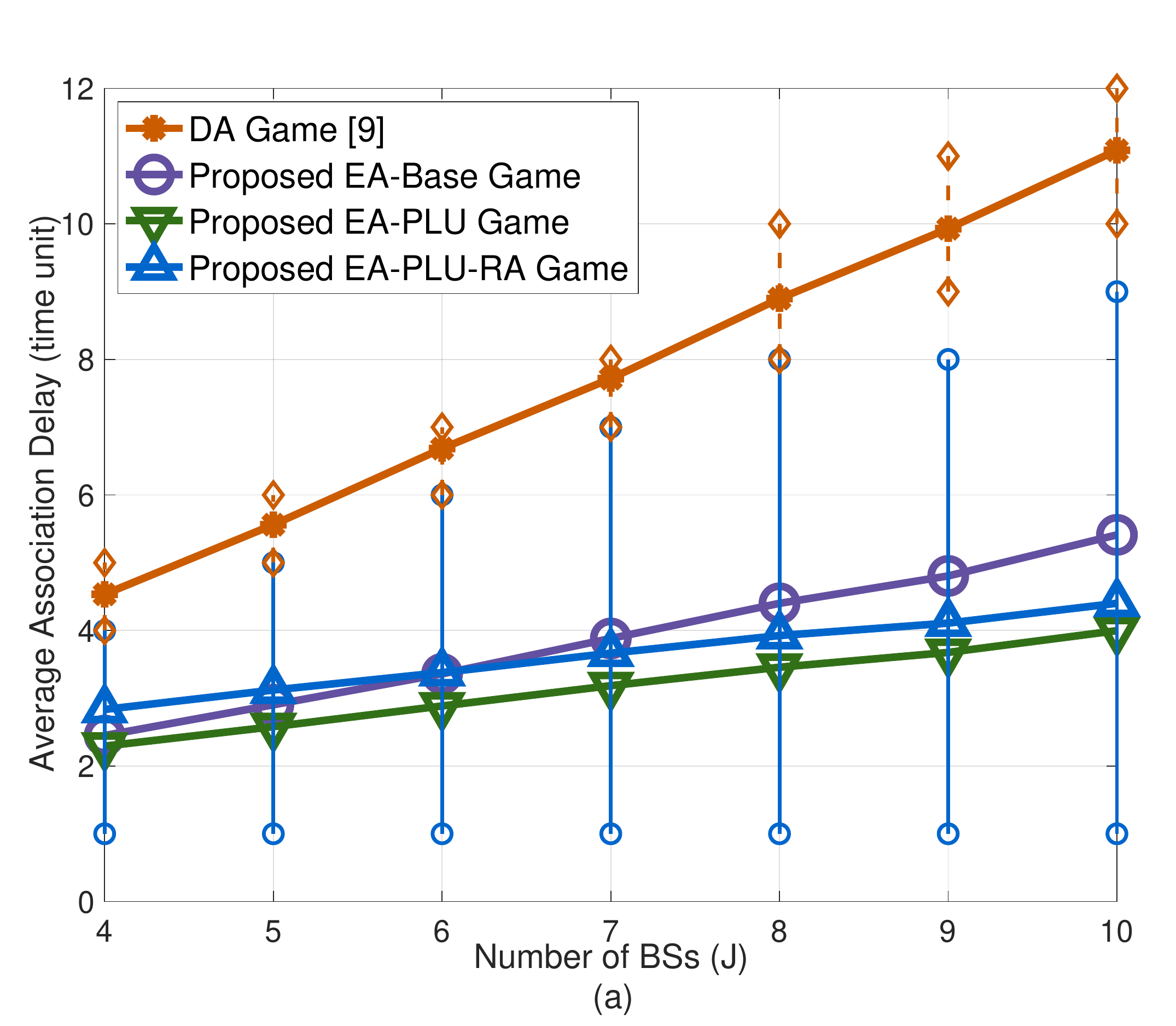}
\hspace*{4em}
\includegraphics[scale=.31]{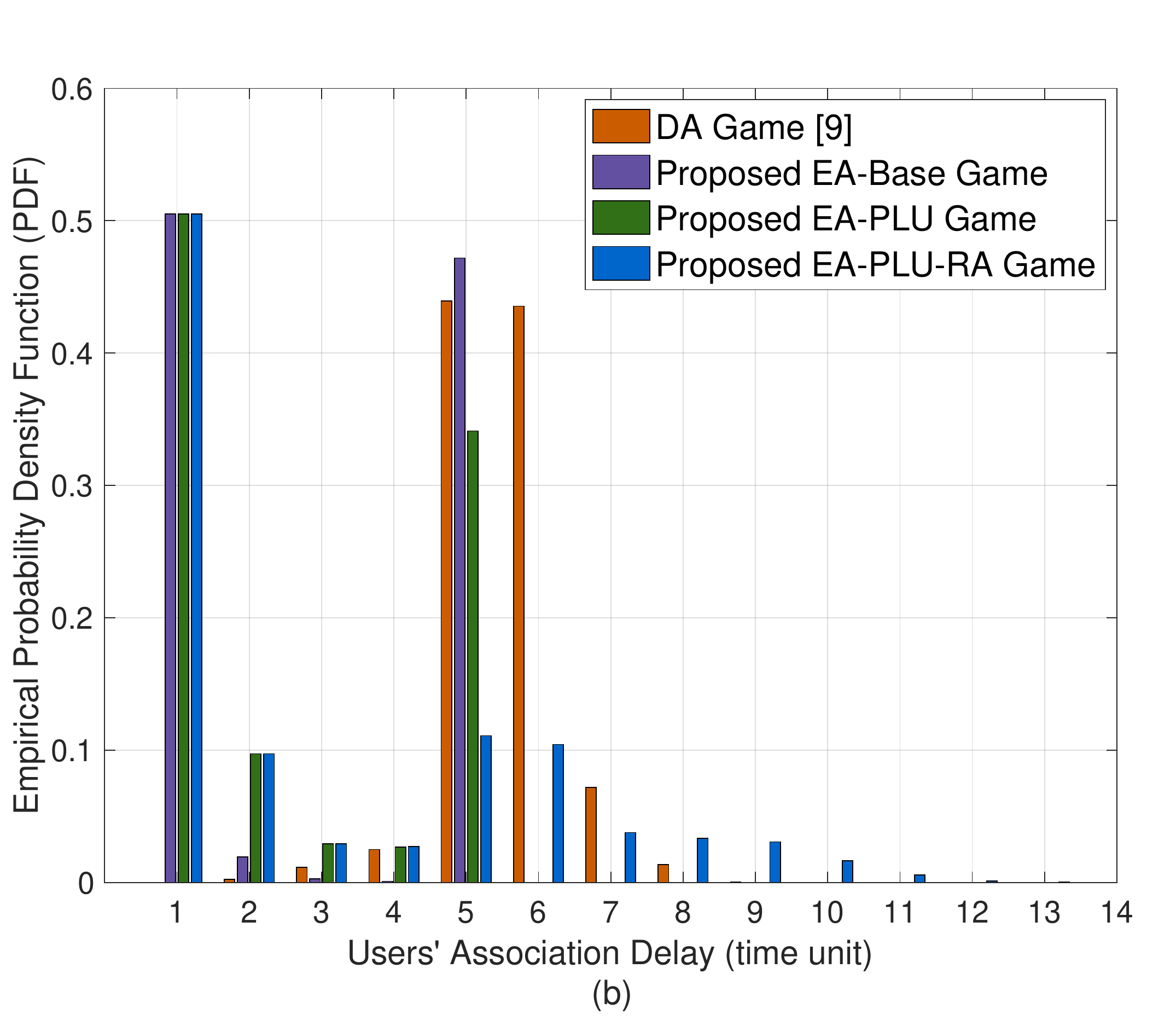}
\vspace*{-0.6em}
\caption{Comparing the association delay of DA and EA games under critical load scenario. a) Average association delay in a HetNet with 1 MSBS and $J-1$ SCBSs, b) Empirical PDF of association delay in a HetNet with $J=5$ BSs and $K=35$ UEs. The vertical lines show the 25 and 75 percentile bars. The time unit is the amount of time for sending an application and receiving a response.}
% Matlab Code: M130_MT_Journal_NumApp_AssDelay.m
% load('M130_MT_Journal_NumApp_AssDelay_50Locs_20Chs_15q_1.mat')
\label{Delay_Appl}
\end{figure*}
\begin{figure*}[t]
\centering
\includegraphics[scale=.31]{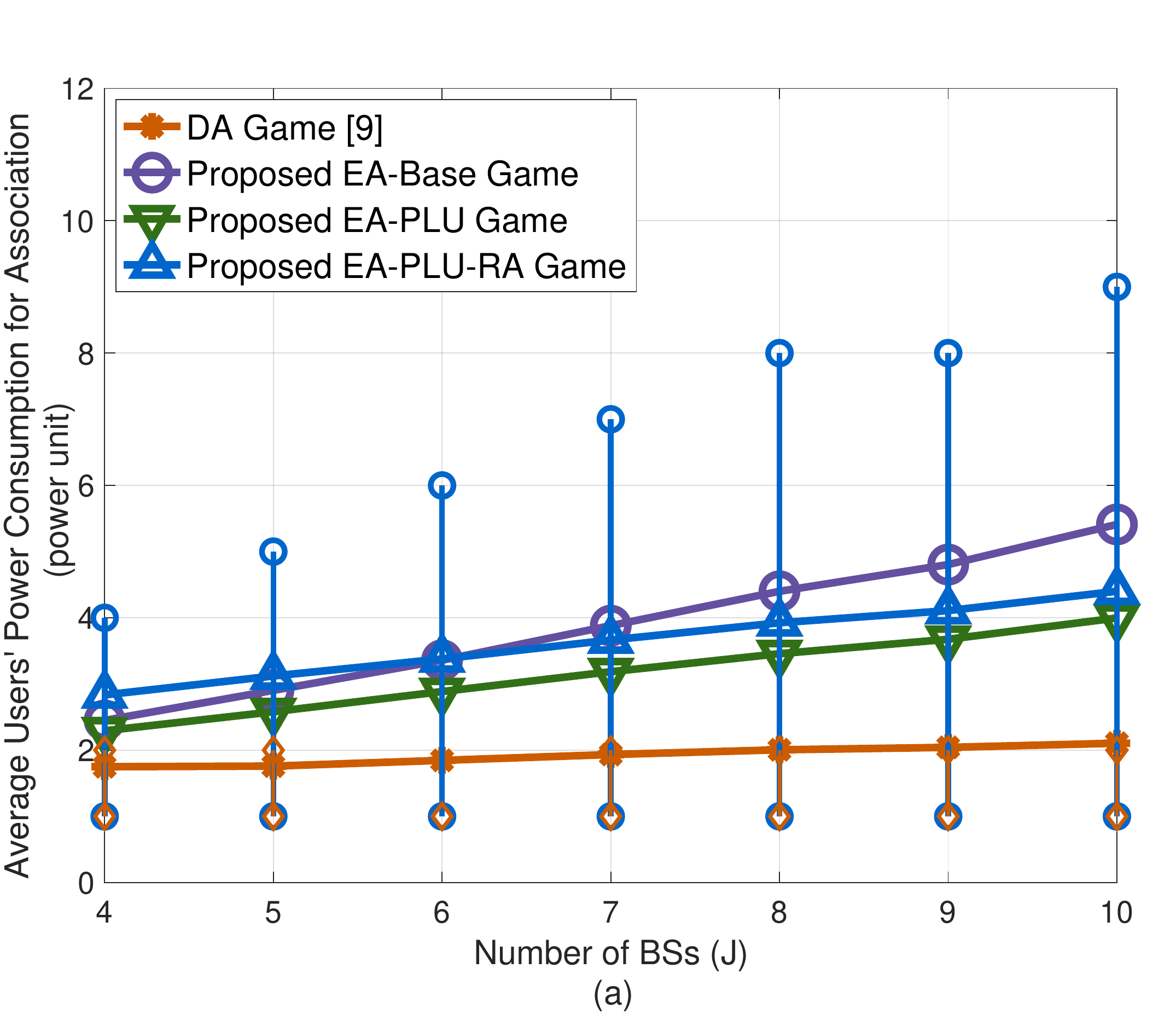}
\hspace*{4em}
\includegraphics[scale=.31]{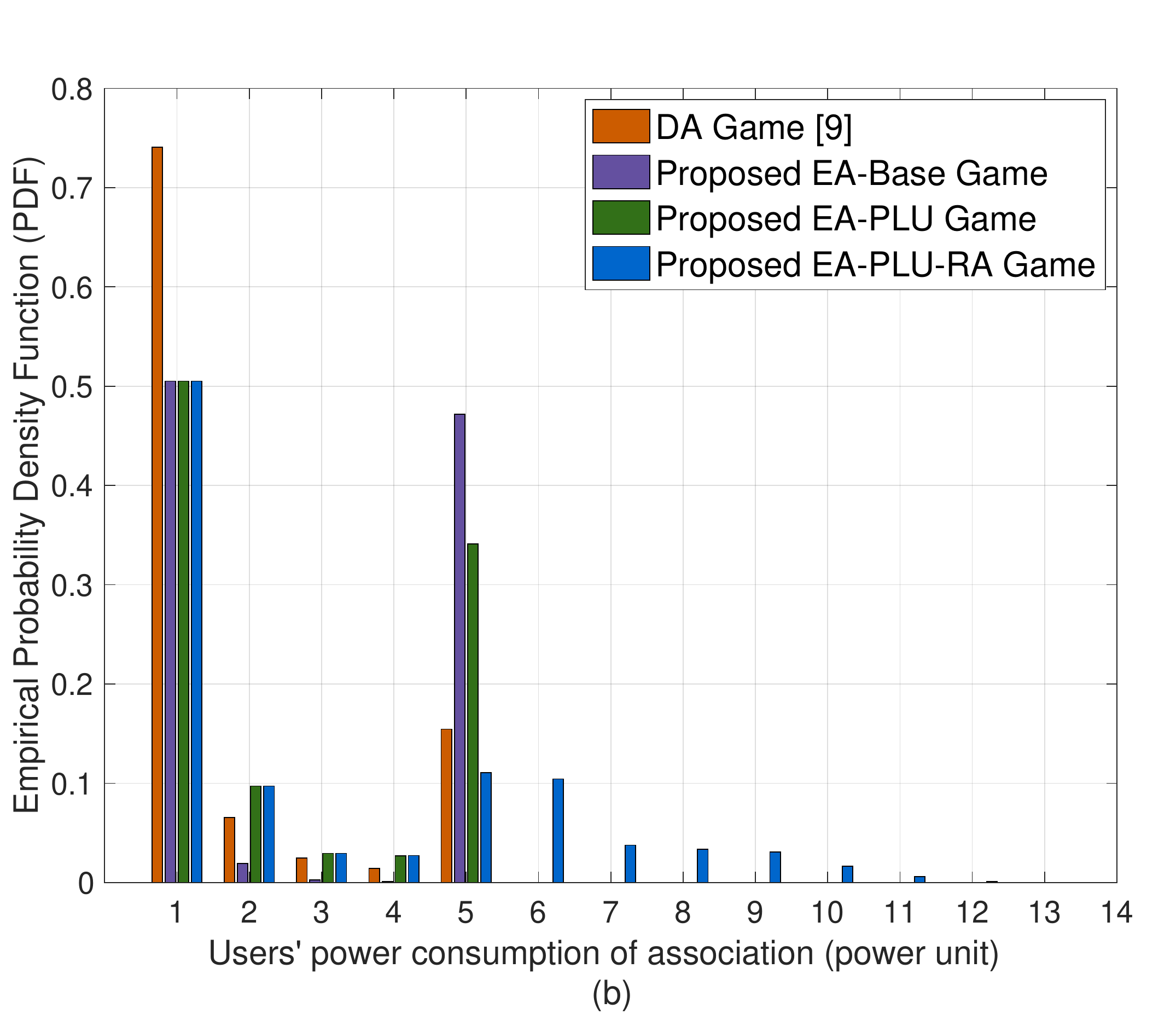}
\vspace*{-0.6em}
\caption{Comparing the power consumption of DA and EA games under critical load scenario with $q_1=15$. a) Average user power consumption for association in a HetNet with 1 MSBS and $J-1$ SCBSs, b) Empirical PDF of user power consumption for association in a HetNet with $J=5$ and $K=35$. The vertical lines show the 25 and 75 percentile bars. The power unit is the amount of power consumed during an application and response step.}
% Matlab Code: M130_MT_Journal_NumApp_AssDelay_PDF_CDF.m
% load('M130_MT_Journal_NumApp_AssDelay_PDF_CDF_100Locs_20Chs_J5_15q_1.mat')
\label{Delay_Appl_PDF}
\end{figure*}

\subsection{Complexity of Multi-Game Matching Algorithm}\label{Comp_MA}
Based on the complexity analysis in Sec. \ref{Comp_MGs}, the complexity of the proposed $N$-game matching algorithm using the DA game is $\mathcal{O}(NJ^2K\log(K))$ and using the EA game is $\mathcal{O}(NJK\log(JK))$, both of which are still much smaller than the cost of centralized WCS algorithm ($\mathcal{O}(M_j^2 K^2\log(K))$) since $N$ and $J$ are usually much smaller than $K$, and $M_j$ is typically a large number. 
We note that all computations in the WCS algorithm are carried out by a central coordinator. In the proposed multi-game matching algorithm, however, computations are distributed among the best-$\mathbold{\beta}$-tracker (for computing the network utility and keeping track of the best association vector), and the BSs and UEs who need to build their own preference lists. A summary of computational complexity of different user association schemes is shown in Table \ref{Comp_UA_schemes}.

\section{Numerical Results}\label{Sim_res}
In this section, we evaluate the performance of the proposed user association matching games and algorithm in the downlink of a mmWave-enabled HetNet with $J$ BSs and $K$ UEs. The network includes 1 MCBS operating at 1.8 GHz with quota $q_1=15$ and $J-1$ SCBSs operating at 28 GHz each with quota $q_j=5, j\in\{2, .., J\}$. Unless otherwise stated, we consider a HetNet with 1 MCBS, 4 SCBSs, and 35 UEs.
Sub-6 GHz channels and mmWave channels are generated as described in Sec. \ref{Ch_Models}. We assume each mmWave channel is composed of 5 clusters with 10 rays per cluster. In order to implement 3D beamforming, each MCBS is equipped with a massive MIMO antenna with 64 elements, each SCBS has a $8\times 8$ UPA, and each UE is equipped with a single-antenna module for sub-6 GHz band, and a 4-element antenna array for mmWave band.
%The noise power spectral density is $-174$ dBm/Hz, and the bandwidths for sub-6 GHz band and mmWave band are 20 MHz and 1 GHz, respectively. 
Also, we assume that the transmit power of MCBS is 10 dB higher than that for SCBSs.
Network nodes are deployed in a $500 \times 500~\textrm{m}^2$ square area where the BSs are placed at specific locations and the UEs are distributed randomly according to a PPP distribution with density $K$ UEs within the given area.

\subsection{Association Delay and Power Consumption}
Fig. \ref{Delay_Appl} compares the DA and EA games in terms of association delay under critical load scenario, with average delay on the left and delay distribution on the right.
Subfigure (a) shows that the EA games significantly outperforms the DA game in terms of average association delay. This advantage becomes more significant as the network size increases.
Subfigure (b) illustrates that all EA games perform better than the DA game in terms of user association delay by having the delay distribution more concentrated around low delay values. For example, the association delay under the EA games for about 50\% of the UEs is only one time unit, while for the DA game, more than 96\% of the UEs have an association delay of at least 5 time units. 
Also, the maximum delay is 8 time units for the DA game, 5 time units for EA-Base and EA-PLU games, and 14 time units for the EA-PLU-RA game. However, it is worth mentioning that the probability of long delay (more than 8 time units) for EA-PLU-RA is only 5\%.

Fig. \ref{Delay_Appl_PDF} compares the matching games in terms of association power consumption under critical load scenario. Subfigure (a) shows that the DA game is slightly more power efficient on the average than the proposed matching games, as it has a lower average user power consumption for association. The additional power consumption of EA games, however, is not substantial and is attributed to the fact that in EA games, each UE may apply on the average more times than in the DA game because of no waiting lists, so that a UE may be rejected and apply again immediately.
According to subfigure (b), about 75\% of UEs only consume one power unit under the DA game, while this value for the EA games is about 51\%. As expected, EA-PLU-RA consumes the most power due to the reapplication process, where the probability of consuming more than the maximum power of other games (5 power units) under EA-PLU-RA is about 23\%. These observations confirm that the EA games results in faster association process, while the DA game is slightly more power efficient. 

\begin{figure}
\centering
  \centering
  \includegraphics[scale=.31]{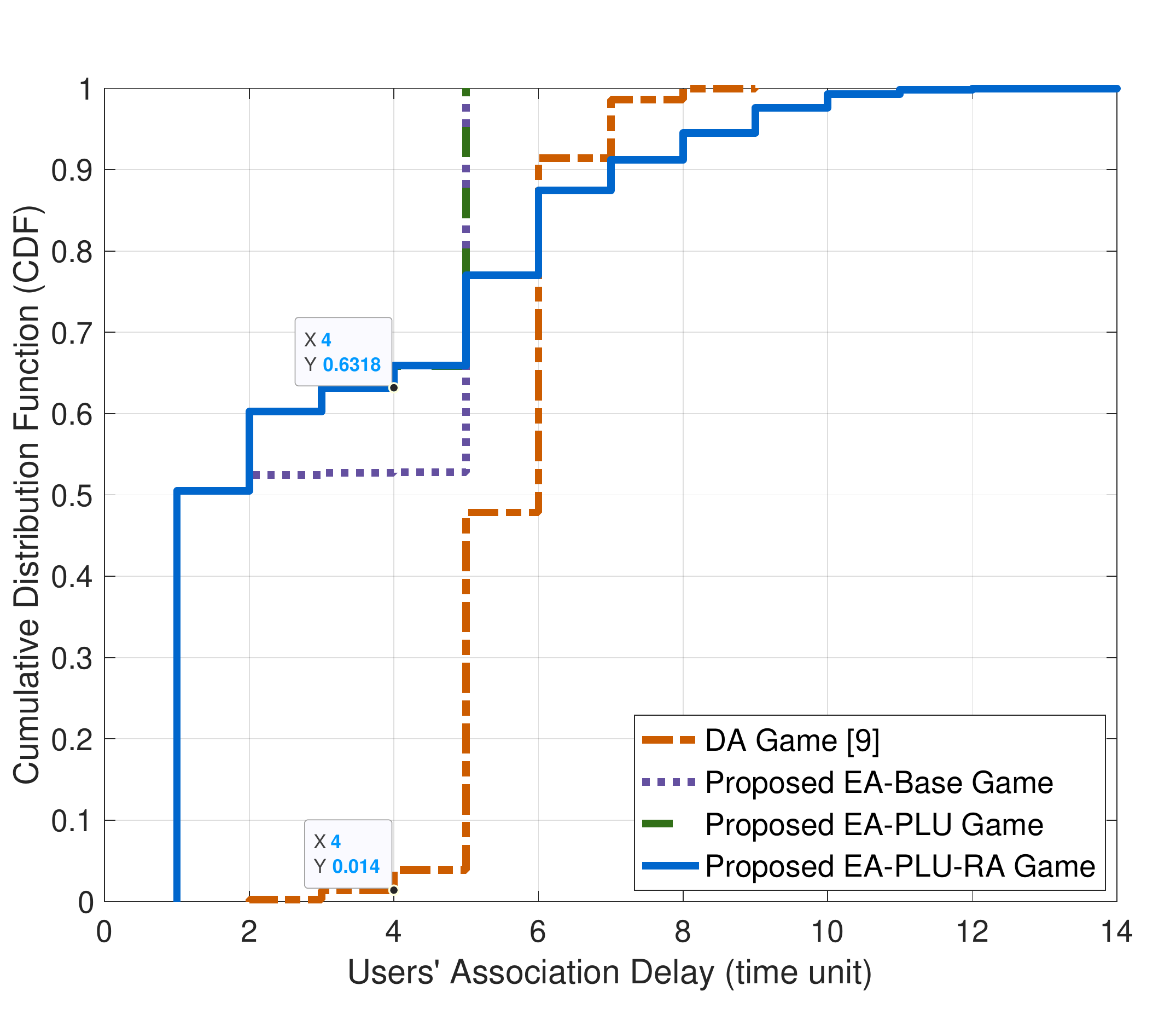}
  \vspace*{-0.5em}
   \caption{Comparing the empirical CDF of user association delay for EA and DA matching games in a HetNet with $J=5$ BSs, $K=35$ UEs. }
   \label{CDF_Delay}
   % M130_MT_Journal_NumApp_AssDelay.m
\end{figure}

Fig. \ref{CDF_Delay} depicts the empirical CDF of user association delay for the user association matching games. This figure again confirms that the probability of having low association delay (below 5 time units) is significantly higher in EA games than in the DA game, and is the highest for the EA-PLU and EA-PLU-RA games. For instance, the probability of having a maximum association delay of 4 time units is 63\% for the EA-PLU-RA game and only 1.4\% for the DA game. These results illustrate that the association process in the EA games is much faster compared to the DA game.

\begin{figure}
  \centering
  \includegraphics[scale=.31]{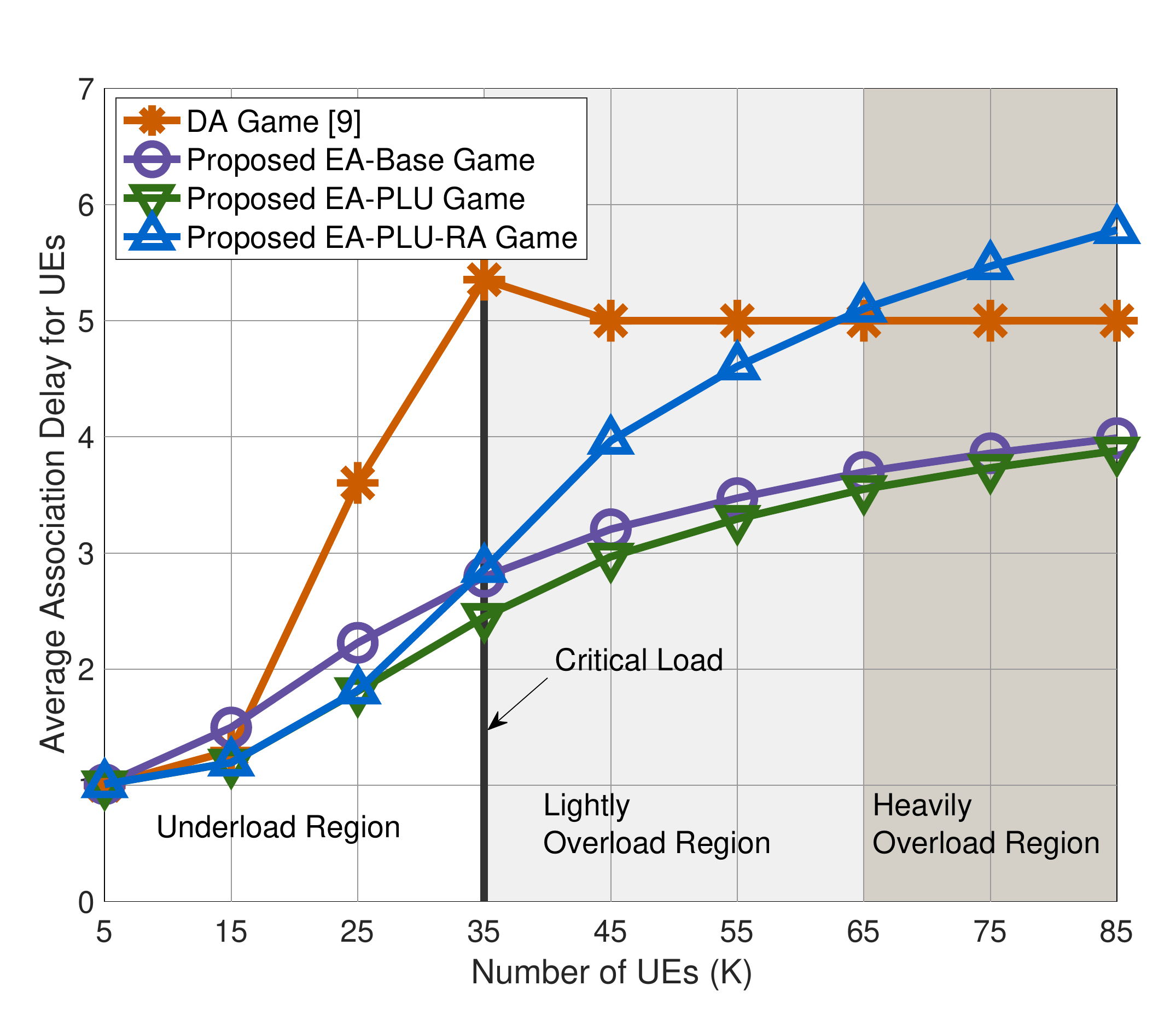}
    \vspace*{-0.5em}
  \caption{Comparing the effect of different loading scenarios on average association delay in a HetNet with $J=5$ BSs and quota vector $\mathbf{q}=[15, 5, 5, 5, 5]$. }
  \label{underload_to_overload}
  % M130_MT_UnderLoad_to_OverLoad.m
% load('M130_MT_UnderLoad_to_OverLoad_100loc_20ch_5BSs_15q_1.mat')
\end{figure}

Fig. \ref{underload_to_overload} depicts the association delay versus the number of UEs in a HetNet with 1 MCBS, 4 SCBSs. Keeping the BSs' quota fixed so that the network can serve a maximum of 35 UEs, we increase the number of UEs such that the network transitions from underload (left shaded region) to critical load (vertical line at $K=35$ UEs), lightly overload (middle shaded region), and finally heavily overload (right shaded region) cases, in order to investigate the effect of different loading scenarios on association delay. We observe an interesting effect that in the DA game, the average delay in the overloading cases are exactly equal to the number of BSs, since the DA game terminates when all UEs are waitlisted by BSs, and this happens at the end of $J$th iteration since each UE has no more than $J$ options. For the EA-Base and EA-PLU games, the association delay is always less than $J$ since UEs do not reapply to BSs. A different trend is observed for the EA-PLU-RA game due to reapplication process and since BSs only accept UEs within their quota. Thus, the higher number of UEs, the more reapplications and the larger the association delay. 
Note that the heavily overload region where the delay of EA-PLU-RA crosses that of DA occurs when the number of UEs in the network is about twice the BSs' total quota, which is unlikely to happen in the real world.

\begin{figure*}[t]
\centering
\includegraphics[width=0.9\textwidth]{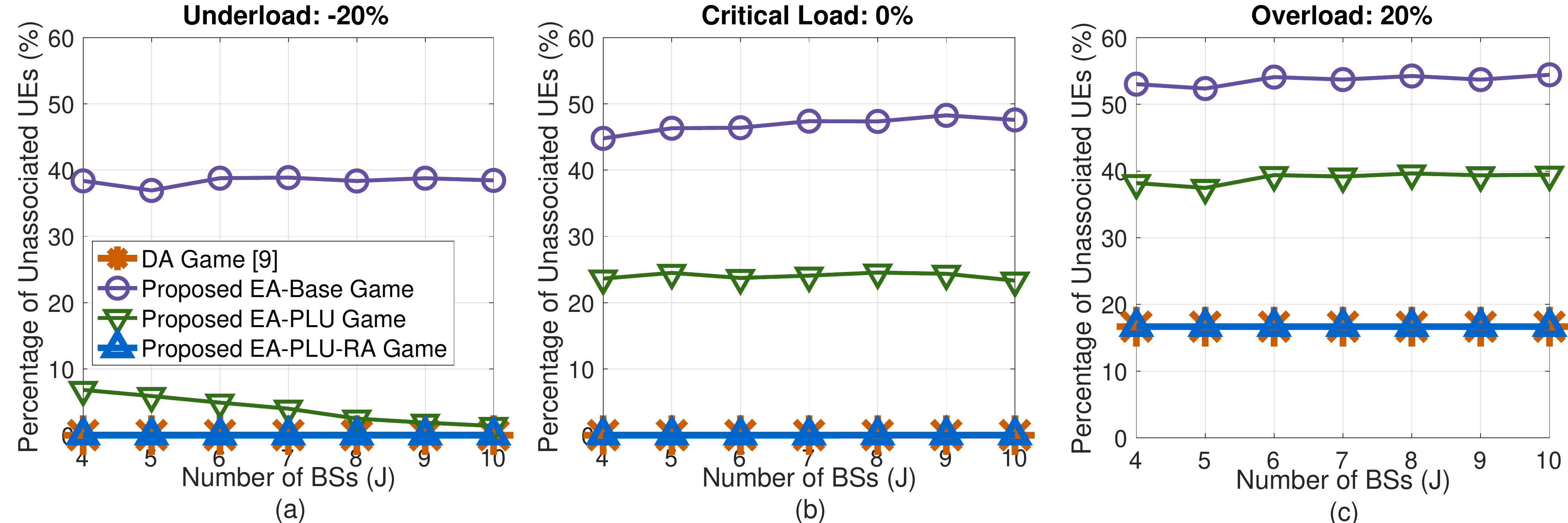}
  \vspace*{-0.5em}
\caption{Percentage of unassociated UEs under three different loading scenarios: a) underload, b) critical load, and c) overload. The number of BSs and UEs increases while BS quotas are fixed $q_1=15, q_j=5, j\in\{2,...,J\}$.}
\label{UnAss}
% New Matlab Codes (for Rev. 1):
% M138_MT_Journal_UnAssUEs_0perc.m
% M138_MT_Journal_UnAssUEs_minus20perc.m
% M138_MT_Journal_UnAssUEs_plus20perc.m
% load('M138_MT_Journal_UnAssUEs_minus20percent.mat')
% load('M138_MT_Journal_UnAssUEs_0percent.mat')
% load('M138_MT_Journal_UnAssUEs_20percent.mat')
% M138_Percentage_of_UnAssUEs_Fig9.m
\end{figure*}

\subsection{Percentage of Unassociated Users}
In practice, the number of UEs in the network is not always equal to the total quota of BSs. Thus, there may be unassociated UEs at the end of an associations game. 
For the next simulation, we increase both the numbers of BSs and UEs while keeping the BSs' quota fixed. 
Fig. \ref{UnAss} compares the percentage of unassociated UEs under three loading scenarios: a) underload, b) critical load, and c) overload.
%, where the percentage on top of each subfigure indicates the extra loading percentage. Specifically, 
For the underload case (a), the number of UEs is 20\% less than the total quota of BSs, whereas for the overloading case (c), the number of UEs is 20\% more. This figure shows that the proposed EA-PLU-RA game has similar performance as the DA game under all three loading scenarios since both games guarantee that maximum number of UEs (limited by BSs’ quotas) are associated at the end of the game (see Lemma 2). Also, it can be inferred that without preference list updating and reapplying steps (EA-Base and EA-PLU), there may be unassociated UEs even for the underload and critical load cases. Thus, these two steps are necessary to make the best use of available resources provided by BSs.

\begin{figure}
\centering
\includegraphics[scale=.31]{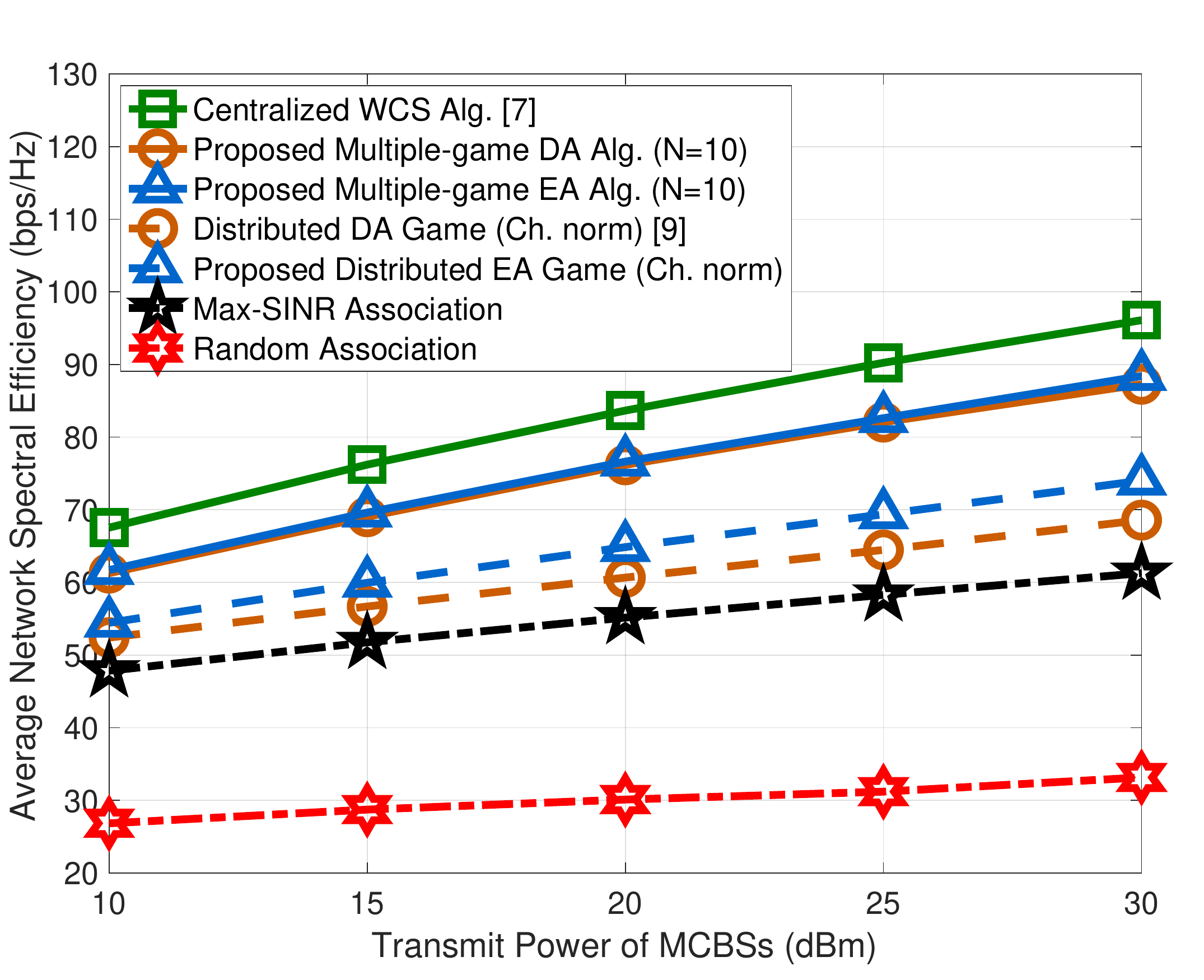}
  \vspace*{-0.5em}
\caption{Comparing the average network spectral efficiency of user association schemes in a HetNet with $J=5$ BSs and $K=35$ UEs.}
\label{SumRate}
% Matlab Code: M138_MT_Journal_NetSpecEff_3GPP_38_901_PL.m
%load('M138_MT_Journal_NetSpecEff_with_Max_SINR_with_UEdrop_500loc_1ch_3GPP_38_901_Pathloss.mat')
\end{figure}

Fig. \ref{SumRate} compares the network spectral efficiency of the HetNet for several association schemes: 1) centralized WCS algorithm \cite{TWC}, 2) proposed multi-game DA algorithm with $N=10$, 3) proposed multi-game EA algorithm with $N=10$, 4) distributed single DA game \cite{gale1962college}, 5) proposed distributed single EA game, 6) max-SINR association, and 7) random association.
The EA game used in this simulation is the one with preference list updating and reapplying (EA-PLU-RA). For the single matching games and initial round of the multi-game algorithms, the preference lists are built based on channel norms which include both instantaneous and large-scale CSI. For the multi-game matching algorithms, we use the matching obtained from the corresponding single matching game as the initial association vectors, and run each algorithm for $N=10$ iterations. At each iteration, the preference lists are updated using the instantaneous rates obtained based on the resulting association vector of that iteration (see (\ref{R_as_PrfList})).
In the max-SINR association scheme, each UE connects to the BS providing the highest received SINR. For random association, each UE randomly associates with a BS based on a uniform distribution. In these two schemes, if a BS is overloaded, the overloading UEs are dropped and become unassociated.

The results show that the EA-PLU-RA game outperforms all other distributed games and approaches the performance of centralized WCS. Interestingly, the EA-PLU-RA game slightly outperforms the DA game  in terms of network spectral efficiency, confirming the fact that stability is a less relevant performance metric for cellular networks.
This higher spectral efficiency is concurrent with lower association delay of the EA game. While this result may appear counter-intuitive at first, it actually makes sense. Although the DA game was shown to be optimal in terms of stability \cite{gale1962college}, it was not known to be optimal for other metrics relevant to cellular systems, including spectral efficiency and delay. Since DA is not optimal for either of these metrics, there is no inherent trade-off between them for the DA game, and as evident by the EA game achieving better performance in both metrics.

This figure shows that compared to the centralized WCS algorithm, multi-game algorithms achieve about 90\% of the performance, and a purely distributed single game can achieve 80\% performance of the centralized algorithm. Such a performance figure is laudable for a distributed algorithm. 
Although max-SINR association has a close performance to matching games in terms of average network spectral efficiency, it results in more unassociated UEs. Random association shows a very poor performance which implies the importance of user association in cellular HetNets.

\begin{figure}
  \centering
\includegraphics[scale=.31]{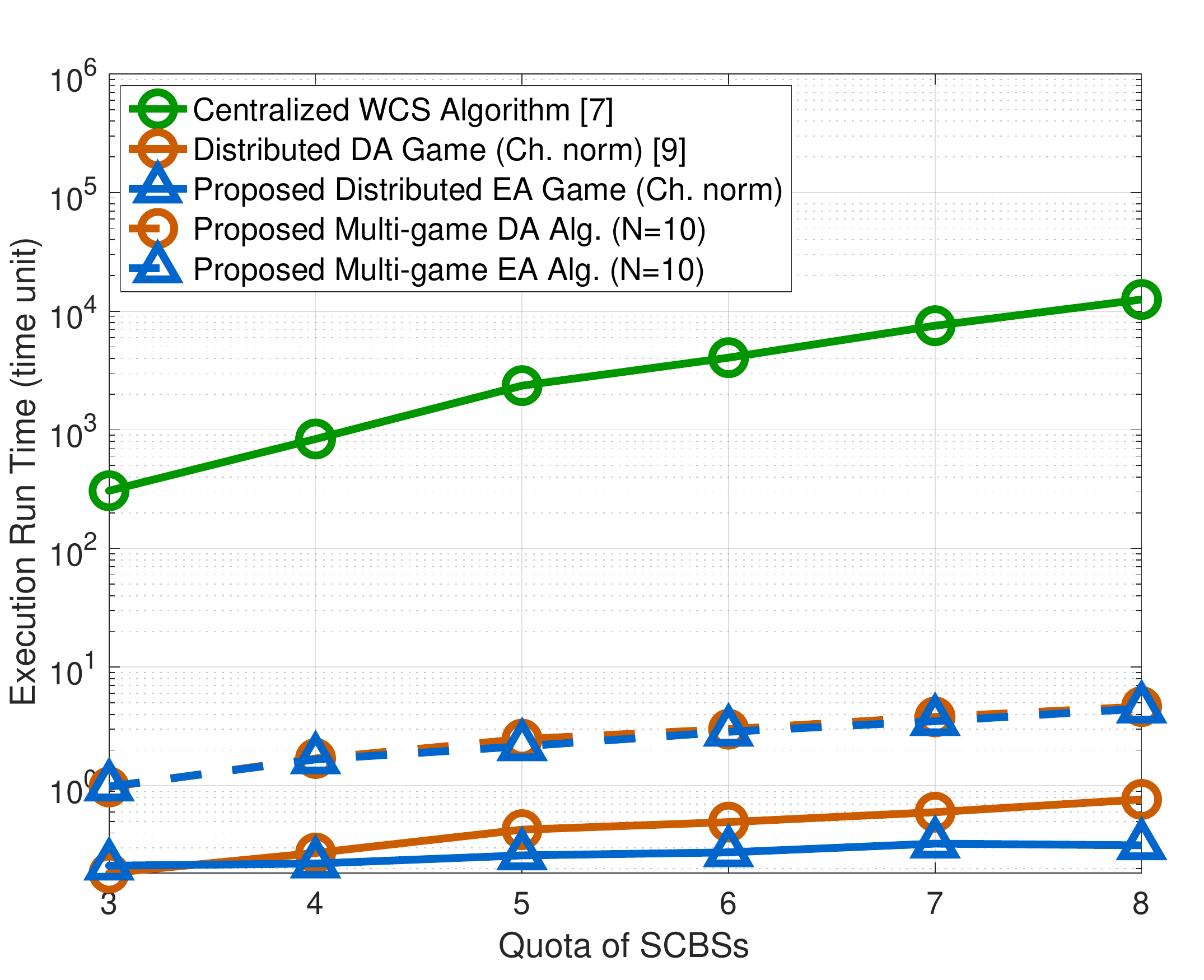}
  \vspace*{-0.5em}
\caption{Comparing the execution run time of user association schemes in a HetNet with $J=5$ BSs and $K=35$ UEs.} 
\label{RunTime}
% M130_MT_RunTime_Increase_Quota.m
\end{figure}

In Fig. \ref{RunTime}, we compare execution run time of the user association matching games and multi-game matching algorithms under a critical load scenario. The vertical axis represents the run time in time unit (which varies depending on capabilities of the processing machine) and the horizontal axis shows the quota of SCBSs. We observe that as the number of UEs increases, the proposed EA-PLU-RA matching game performs faster than the DA game. The trend is similar for the multi-game matching algorithms. We also observed there is a stark difference between the run times of centralized WCS algorithm and distributed matching games/algorithms. This result validates the advantage of distributed matching games as the matching games/algorithms have much lower complexity than the centralized WCS algorithm.

\section{Conclusion}
We proposed a set of distributed early acceptance (EA) user association matching games for 5G and beyond HetNets, and compared their performance with the well-known DA matching game. 
We showed that at slightly more power overhead, the EA games result in a significantly faster association process compared to the DA game while achieving better network spectral efficiency. 
The EA-PLU-RA game with preference list updating and reapplication provides the best overall network performance in terms of both association delay and percentage of associated users. 
These results suggest that stability is a less relevant metric for user association and EA may be more suitable in B5G wireless networks for real-time distributed association.

Next, we proposed a multi-game matching algorithm to further enhance the network spectral efficiency by running multiple rounds of a matching game. Numerical results show that the proposed distributed EA games and multi-game algorithm achieve a network spectral efficiency within 80-90\% of the near-optimal centralized WCS benchmark algorithm, while incurring a complexity at several orders of magnitude lower and significantly less overheads due to their distributed or semi-distributed nature. 

\ifCLASSOPTIONcaptionsoff
  \newpage
\fi

\bibliographystyle{IEEEtran}
\bibliography{References}

\end{document}